\documentclass[12pt]{report}
\usepackage{graphicx}
\usepackage{subcaption}
\usepackage{float}
\usepackage{booktabs} % 在导言区引入booktabs包
\usepackage[english]{babel}
\usepackage{makecell} %
\usepackage{pgf-pie}
\usepackage{tikz}
\usepackage[letterpaper,top=2cm,bottom=2cm,left=3cm,right=3cm,marginparwidth=1.75cm]{geometry}
\usepackage{hhline} % Add this to your LaTeX preamble for double lines
\usepackage{amsmath}
\usepackage{graphicx}
\usepackage{booktabs}
\usepackage{caption}
\usepackage[colorlinks=true, allcolors=blue]{hyperref}
\usepackage{pdfpages}
\usepackage{setspace}
\begin{document}
% \maketitle
% \begin{titlepage}
% \centering
% \vspace*{2cm} % 页面顶部添加空间，移动图片和文字
% \begin{center}
%   \includegraphics[width=0.2\textwidth]{uqlogo.png} % 图片大小和文件名按需调整
%   \\[1em] % 该命令在图片和标题之间添加了额外的空间，按需调整
%   AUSTRALIA
% \end{center}

% {\Large \textbf{THE UNIVERSITY OF QUEENSLAND}}\\[2cm] % 
% \vspace{-5em}
% {\Large \textbf{School of Information Technology and Electrical
% Engineering}}\\[1cm] % 标题

% {\Large \textbf{Security study based on the CHATGPT plugin system:\\Identifying Security Vulnerabilities in the ChatGPT Plugin System (Project Proposal)}}\\[0.5cm] % 标题

% {\large Ruomai Ren 47652749}\\
% {\large April 17, 2024}\\
% \vspace*{\fill}
% \end{titlepage}

\newpage
\includepdf[pages=-]{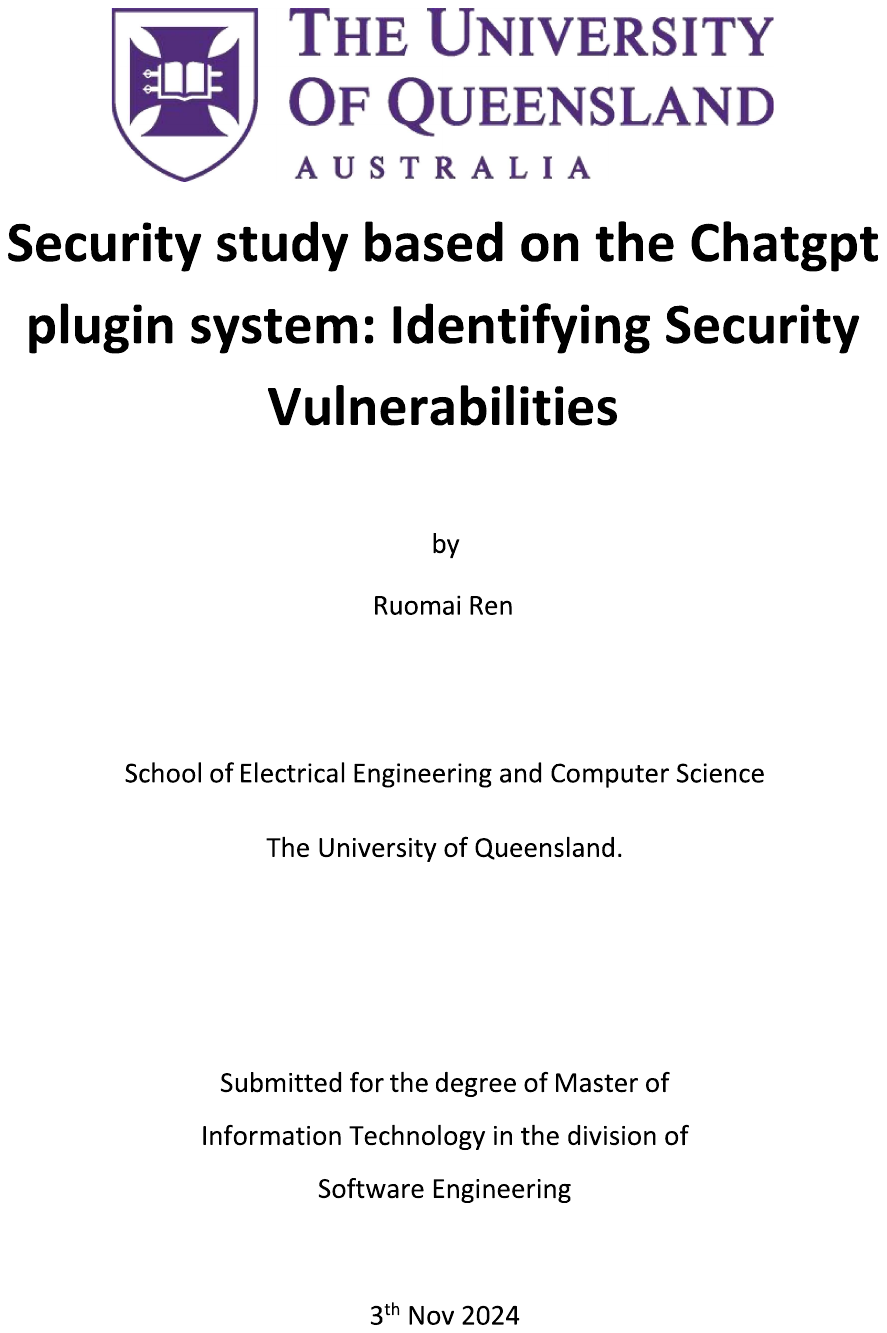}
\addcontentsline{toc}{chapter}{Acknowledgements} % 将添加到目录中

\pagenumbering{roman} % 使用罗马数字作为页码

% 摘要标题和内容
\chapter*{Abstract} % 使用无编号章节
Plugin systems are a class of external programmes that provide users with a wide range of functionality, and while they enhance the user experience, their security is always a challenge. Especially due to the diversity and complexity of developers, many plugin systems lack adequate regulation. As ChatGPT has become a popular large-scale language modelling platform, its plugin system is also gradually developing, and the open platform provides creators with the opportunity to upload plugins covering a wide range of application scenarios. However, current research and discussions mostly focus on the security issues of the ChatGPT model itself, while ignoring the possible security risks posed by the plugin system. This study aims to analyse the security of plugins in the ChatGPT plugin shop, reveal its major security vulnerabilities, and propose corresponding improvements. Notably, some of the findings from this report have subsequently been incorporated into academic publications~\cite{10.1145/3691620.3695510}.

\addcontentsline{toc}{chapter}{Abstract} % 将摘要添加到目录中

% \begin{titlepage}
%     \vspace*{1in} % 控制标题顶部的空白距离
%     \noindent\textbf{\Large Abstract} % 左对齐标题并设置粗体和字号
%     \vspace{1em} % 标题和正文之间的空白
    
%     \begin{spacing}{1.5} % 设置行距为1.5倍
        
%     \end{spacing}
    
    % \vfill % 将内容垂直居中对齐
% \end{titlepage}

\pagenumbering{arabic} % 从摘要之后使用阿拉伯数字作为页码

% 目录
\tableofcontents
\listoffigures   % 生成插图目录
\listoftables    % 生成表格目录
\newpage

%基于周四完成的---文本写完，加不同商店比较
% 多写文献综述

% 查ai率，
% 查重，
% 检查内容逻辑，

% 加图，
% 加不同商店比较

%格式加完--前面乱起八糟的东西
%字数够的话，评估单独写，鲁棒性之类的问题

%引用，外部文件提交

%15页
% \begin{abstract}

\chapter{Introduction}

Plugin systems, external programs that enhance the user experience through a range of features. However, the security of extensions and plugin systems always poses a number of challenges, and given the complexity and diversity of their developers, they are often not fully regulated. For example, in the case of browser plugins, Eriksson's empirical study in 2022 showed that there are still nearly a thousand browser extensions that are subject to varying degrees and multiple means of security threats and contamination\cite{ErikssonBenjamin2022Htsa}.

        In the past year, CHATGPT has become recognized as the hottest LLM platform, and its breadth of application and prevalence has gained public attention. Among them, CHATGPT plugin is a plugin application store launched by OPENAI, where creators freely upload plugins for users to use, and as of now there are more than 1,000 of them , and the plugins cover a variety of uses including PDF processing, voice processing, planning airplane flights, and so on\cite{chatplugin}.
        
        However, chatgpt as a commercialized large language model, its security is also widely discussed, although itself has good security, but the security of the extension system is often ignored. Currently, when discussing the security of chatgpt platform, it mainly focuses on the tampering and generative harmful information of the nlp model itself\cite{KangDaniel2023EPBo}\cite{LiHaoran2023MJPA}, ignoring the vulnerability and security holes as the extension program itself. In fact, extensions, as external applications that can invoke the NLP model, have not received much security attention.
        The aim of this paper is to explore the security of plugins in the CHATGPT plugin store, reveal the main security vulnerabilities that exist, and suggest improvements.

\chapter{Background and Literature Review}

\section{Current State of Security in Plugin Systems and Chatgpt}

% 插件系统在现代应用中得到广泛应用，能够扩展平台的功能并提升用户体验。然而，随着插件数量的增加，安全性问题也日益凸显，不当的权限管理和数据泄露成为主要风险。

% 本节将首先回顾当前 ChatGPT 插件系统的安全讨论方向，以及其他大型插件系统在安全性方面的现状。随后，我们将对不同插件系统的关键特性和安全管理进行比较分析，深入探讨插件的关键架构和组件，最后总结当前插件系统的主要安全问题，并阐述本研究的动机。
Plugin systems are widely used in modern applications to extend the functionality of the platform and enhance the user experience. However, as the number of plugins increases, security concerns are becoming increasingly apparent.

In this section, we will first review the direction of the current security discussion of ChatGPT and the current status of other large-scale plugin systems in terms of security. 

Subsequently, we will provide a analysis of different plugin system features and security management, and finally summarise the possible security issues of current plugin systems and describe the motivation of this research.

\subsection{The Current Direction of CHATGPT's Security Discussion}

Recent research on the security of LLM platforms, mobile operating systems, and various ecosystems has systematically explored risks such as third-party dependencies and knowledge leakage in the GPT ecosystem, privacy policy changes, and data compliance, highlighting the urgent need for cross-domain security and governance in AI ecosystems\cite{yan2025trackinggptspartyservice,10.1145/3696410.3714755,10.1145/3660826,10.1145/3597503.3639107,yan2024quality}.

% 同时，大型应用商店往往难以做到全面的审查，比如，根据Sheryl Hsu等人的研究，CWS的审核机制存在“漏斗效应”，难以全面审查所有插件，尤其在一些插件具备隐蔽行为的情况下，审核和维护变得更加困难,插件商店无法完全保障所有插件的安全.
% 根据Asmit Nayak等人研究，大型浏览器扩展未经授权访问敏感权限和数据的情况非常普遍，且缺乏有效的审核和监管。这导致用户隐私和数据安全面临显著风险
Many studies have shown that ChatGPT performs well in many complex activities, reflecting the breadth of its application areas. For example, the research of Kazama and Shuzo\cite{kazama2024chatgpt} pointed out that ChatGPT can help machine learning beginners to successfully build a highly accurate human activity recognition system through dialogue support and basic model integration in the absence of expert guidance. The research of Li\cite{li2024hot} et al. demonstrated the application potential of ChatGPT in detecting and distinguishing hate, offensive, and toxic (HOT) comments in social media, providing practical value for large-scale content review. The ChatBR system proposed by Bo et al\cite{bo2024chatbr}. uses ChatGPT to automatically evaluate and improve the quality of error reports, and significantly improves the completeness and accuracy of reports by supplementing missing information. Zhang et al.\cite{zhang2024prompt} improved the effect of ChatGPT in software vulnerability detection by optimizing the prompt design, proving the role of structured prompts in improving detection accuracy. The research of Mohajer et al.\cite{mohajer2024effectiveness} further showed that ChatGPT performed well in static analysis tasks (such as null pointer dereference and resource leak detection), demonstrating its potential in error detection and eliminating false positives.

Baesd on that, the current discussion of security issues in the mainstream of CHATGPT focuses on the following areas, which are not only technological innovations, but also involve ethical and legal issues at the same time:

Content policy bypass: Jailbreaking, adversarial questioning, and other means of bypassing CHATGPT's content policies in order to obtain information or responses that violate those policies; kinds of behavior that can lead to the generation of harmful content, such as hate speech, false information, or other inappropriate content\cite{GuptaMaanak2023FCtT}.Besides, ChatGPT often generates false citations in its answers, with only 14\% of the citations being real and supporting the answers, revealing issues with the model’s reliability in citing evidence\cite{zuccon2023hallucinates}.

Instant generation of attack code and phishing text: CHATGPT is capable of generating text on-the-fly based on user input, and the potential for abuse of this feature includes the generation of code for malware, attack scripts, or bait text used in phishing attacks\cite{Al-HawawrehMuna2023Cfcp}\cite{2024Aneo}.Given the nature of the LLM model itself, the content generated by this type of exploitation can be very persuasive and targeted, with a considerable potential to increase the success rate of cyber attacks.

Another popular direction is to use chatgpt to generate content to assist in code review, detect code odour, mitigate code quality issues, reduce developer workload and improve review efficiency.\cite{watanabe2024chatgptcodereview}\cite{silva2024codesmells}\cite{liu2024refining}.

Recently, some studies have started to focus on emerging risks in the GPT application ecosystem. For example, Yan et al.\ proposed an automated approach for tracking GPTs' third-party services, providing in-depth analysis and insights into service dependencies~\cite{yan2025trackinggptspartyservice}. Additionally, further work has been conducted to understand and detect file knowledge leakage within the GPT app ecosystem, offering new perspectives for the prevention and mitigation of knowledge leakage risks in large language model applications~\cite{10.1145/3696410.3714755}.

% 研究表明，ChatGPT在多项复杂活动中表现优异，体现了其应用领域的广泛性。Kazama和Shuzo的研究指出，ChatGPT能够帮助机器学习初学者在缺少专家指导的情况下，通过对话支持和基础模型集成，成功构建出高准确度的人类活动识别系统。Li等人的研究展示了ChatGPT在检测和区分社交媒体中的仇恨、冒犯和有毒（HOT）评论方面的应用潜力，为大规模内容审核提供了实际价值。Bo等人提出的ChatBR系统利用ChatGPT自动评估和改进错误报告的质量，通过补充缺失信息显著提高了报告的完整性和准确性。Zhang等人则通过优化提示设计，提升了ChatGPT在软件漏洞检测中的效果，证明了结构化提示对检测准确度的提升作用。Mohajer等人的研究进一步显示，ChatGPT在静态分析任务（如空指针解引用和资源泄漏检测）中的表现优异，展示出其在错误检测和消除误报方面的潜力

\subsection{The State of Security in Other Large Plugin Systems}
In exploring the security of large plugin systems, we find that these systems exhibit complexity and diversity\cite{wangzihan1,wangzihan2,wangzihan3}. This complexity is mainly manifested in the openness and extensibility of the plugin system, which promotes the richness and personalization of the functions, but also to the plugin system with a large sample size\cite{yan2024privacy_android}. As the plugin system with a large sample size, chrome and firebox, for example, the security problem of the plugin system, the common attack directions are password stealing, traffic stealing, inter-extension attack, cross-site scripting attack and so on\cite{PerdisciRoberto2019WEDa}\cite{KirkJeremy2014MCbe}\cite{VarshneyGaurav2018MbeA}.

In recent years, with the development of technology, a new type of attack means, including the use of computer graphics to bypass the traditional text information protection measures\cite{ErikssonBenjamin2022Htsa}\cite{VarshneyGaurav2018MbeA}. Instead of extracting information directly from the text input, attackers identify passwords and usernames by obtaining screenshots. The danger of this method is that screenshot permissions are relatively easy to obtain and not easily detected by traditional security measures.

In both browsers, Chrome provides an isolated execution environment for each plugin through its "sandbox" mechanism, which restricts the plugin's ability to access user data and other plugins\cite{ErikssonBenjamin2022Htsa}. This isolation measure effectively reduces the possibility of inter-plugin attacks and restricts the access of malicious plugins to user data.

In contrast, Firefox has more limited measures for security isolation, which makes it a greater challenge for plugin security\cite{ErikssonBenjamin2022Htsa}. While Firefox offers a rigorous review process and a range of security policies to protect users from repercussions, the lack of isolation mechanisms similar to Chrome's means that once malicious plugins are installed, they are more likely to affect the rest of the browser and the security of users' data.

According to Asmit Nayak et al\cite{nayak2024experimental}, unauthorised access to sensitive permissions and data by large browser extensions is widespread and lacks effective auditing and regulation. This leads to significant risks to user privacy and data security, and the means to do so are varied and evolving\cite{starov2017extended}.
At the same time, it is often difficult for large-scale app shops to conduct a comprehensive review. For example, according to Sheryl Hsu et al.'s study\cite{hsu2024chromewebstore}, the review mechanism of CWS has a “funnel effect”, which makes it difficult to review all plugins in a comprehensive way, and it is even more difficult to review and maintain the review, especially in the case of some plugins that have hidden behaviours, and the plugin shops are unable to completely guarantee the security of all plugins. The plugin store is not able to fully guarantee the security of all plugins.

\subsection{Comparison of Plugin Store across Multiple Dimensions}
%为了对插件市场安全现状有更好的了解，我们选取了不同维度的插件商店进行比较，其中：

% Chrome 和 Firefox 是全球主流的浏览器扩展商店，分别由 Google 和 Mozilla 运营。它们管理着上万种扩展插件，用户可以通过这些插件来增强浏览器的功能。这些扩展直接在用户的本地浏览器中运行，同时带来了权限管理和数据安全的挑战。和 ChatGPT 插件类似，这两大扩展商店展示了插件在功能上的广泛应用。

% Notion 是一种云端协作和知识管理工具，提供 API 集成功能，允许用户和开发者通过远程调用访问和操作数据。Notion API 用于扩展notion功能，比如，将其和googledriver进行有机结合和数据同步。他通过 OAuth 2.0 授权机制保护用户数据。和 ChatGPT 相似，Notion 也是一种基于线上第三方 API 进行插件和用户沟通，体现出灵活得特点。

% 我们通过比较发现：
% Notion API 集成：在第三方服务器上在线远程执行，依赖 API 进行数据交互，将数据在远程服务器中处理后返回给用户。通过 OAuth 2.0 等认证方式控制访问权限，用户可以授权特定页面或数据库的访问。主要适用于数据同步、项目管理、任务自动化和跨平台数据共享等场景。技术灵活，但场景较为局限。

% Firefox 扩展：在用户的本地浏览器中直接执行，通过浏览器权限管理机制控制访问，用户在安装前需确认权限，安装后可调整权限。支持直接修改或扩展浏览器功能，操控网页内容。插件种类繁多，数量巨大，开发者上传整个应用程序代码来发布插件，不可以在线更新。

% Chrome 扩展：在用户的本地浏览器中直接执行，权限管理方式与 Firefox 类似，安装前确认权限，安装后用户可以查看和调整权限。插件种类繁多，数量巨大，开发者上传整个应用程序代码来发布插件，不可以在线更新。

% ChatGPT 插件：在第三方服务器上在线远程执行，通过 API 调用外部服务，实时在对话中提供数据支持。用户可以在会话中选择启用特定插件，权限仅限于会话内访问。插件种类繁多，数量尚可。同时交互过程中受llm大模型影响
% 在这些比较中可以发现，ChatGPT 插件作为一种新型插件系统，具备以下三个区别于其他的特点：
%其一是依托于强大的 LLM（大型语言模型），极大地扩展了插件的灵活性；
%其二是支持广泛的插件功能，覆盖了多种需求，对权限的要求多样化；
%其三是采用了激进而灵活的线上 API 交互，使用户在对话中实时调用 API，进行多样化的交互。插件更新相对灵活

%虽然这种高灵活性和实时 API 调用使其在多场景表现突出，也意味着潜在的安全和隐私风险更为复杂。与传统插件相比，ChatGPT 插件需要更严格的管理策略和安全规范来保护用户隐私和数据。同时，还要评估llm对其造成得潜在影响
To gain a better understanding of the current state of security in the plugin market, we have selected several dimensions of plugin shops for comparison, including:

\begin{figure}[ht]
    \centering
    \begin{minipage}{0.32\textwidth}
        \centering
        \includegraphics[width=\textwidth]{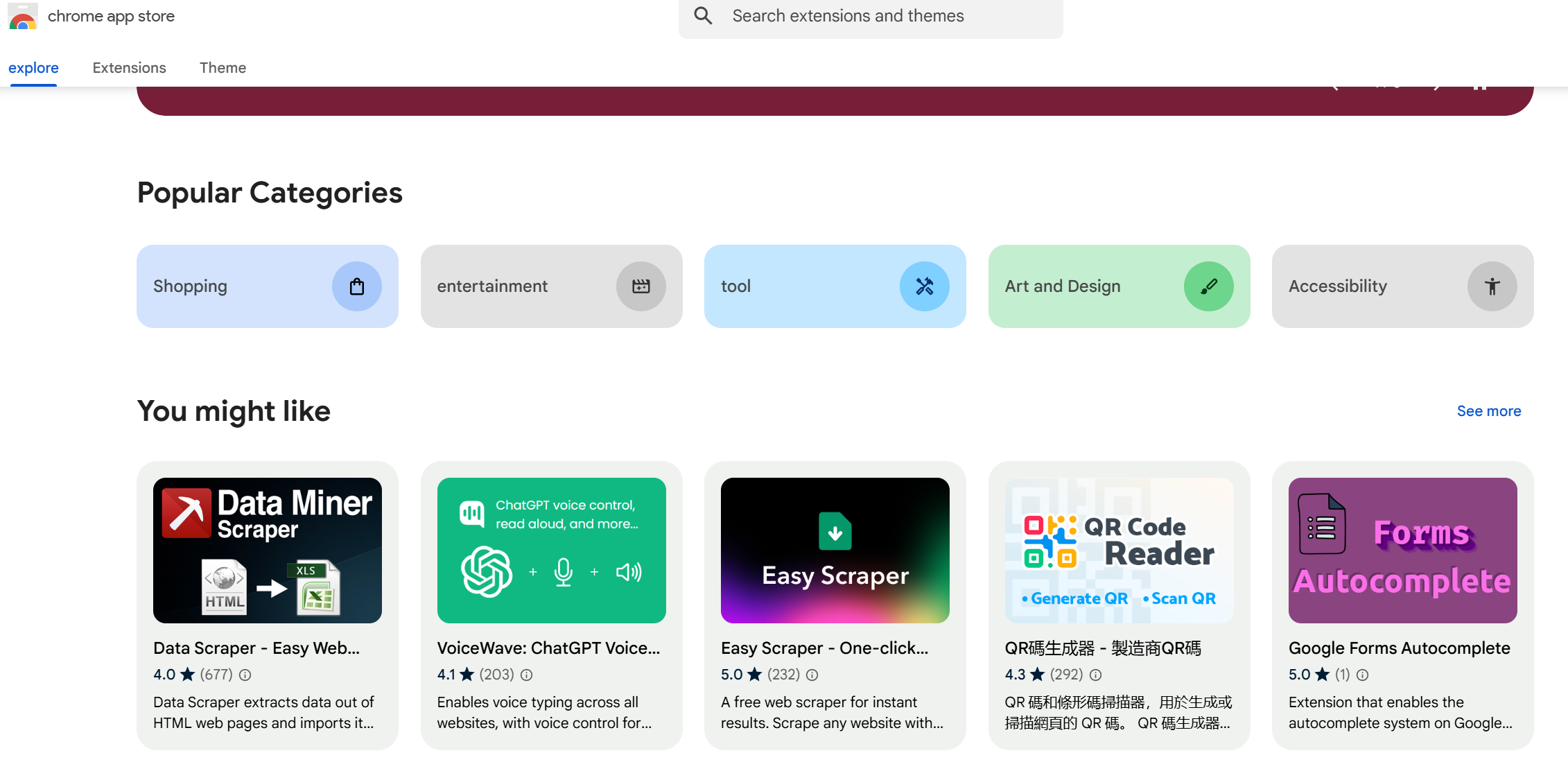}
        \caption{Chrome}
    \end{minipage}
    \hfill
    \begin{minipage}{0.32\textwidth}
        \centering
        \includegraphics[width=\textwidth]{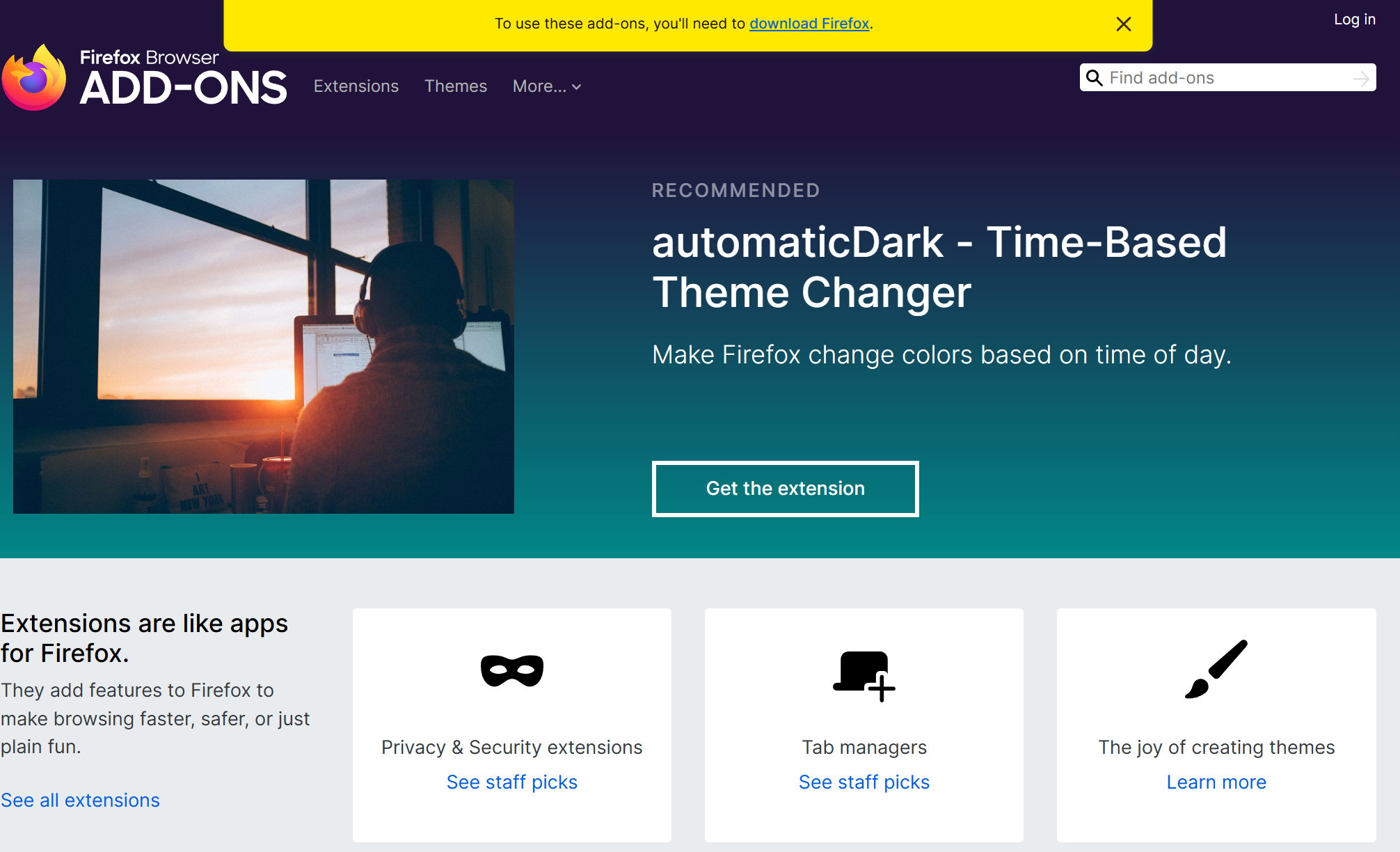}
        \caption{Firefox}
    \end{minipage}
    \hfill
    \begin{minipage}{0.32\textwidth}
        \centering
        \includegraphics[width=\textwidth]{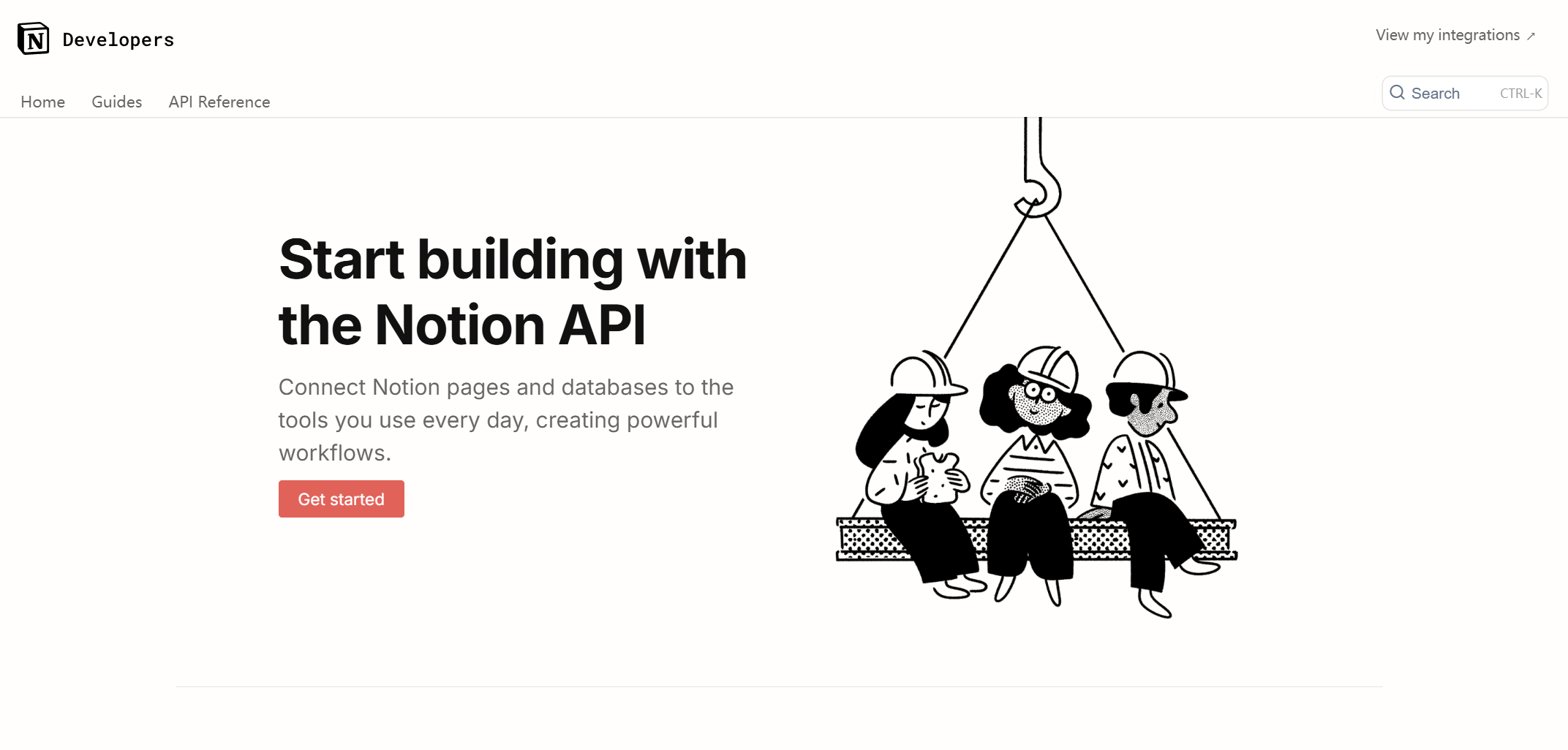}
        \caption{Notion}
    \end{minipage}
    \caption{Comparison of Three Plugin Platforms}
    \label{fig:three_images}
\end{figure}

\begin{itemize}
    \item \textbf{Chrome and Firefox Extensions}: These are the world’s leading browser extension platforms, operated by Google and Mozilla, respectively. They manage tens of thousands of extensions that allow users to enhance their browser functionality. Since these extensions run directly in the user's local browser, they pose unique challenges in terms of permissions management and data security. Similar to the ChatGPT plugin, these two extension platforms demonstrate the wide range of functionality that plugins can offer.

    \item \textbf{Notion}: Notion is a cloud-based collaboration and knowledge management tool that provides API integration, allowing users and developers to access and manipulate data via remote API calls. For example, the Notion API enables users to extend Notion's functionality by integrating with other services, such as Google Drive, for data synchronization. The Notion API protects user data through OAuth 2.0 authorization. Similar to ChatGPT, Notion also relies on third-party APIs, providing flexibility for various integrations.
\end{itemize}

\subsubsection{Findings of Four Plugin Systems}

Through this comparison, we have identified key characteristics of each plugin system:

\begin{itemize}
    \item \textit{Notion API Integration\cite{notionAPI}}: 
    
        It is executed remotely on a third-party server and relies on API calls for data interaction and processing.
        After the data is processed, it is returned to the user. And OAuth 2.0 and other authentication methods are used to manage access rights, allowing users to authorize access to specific pages or databases.
        It is mainly suitable for data synchronization, project management, task automation, and cross-platform data sharing. It is technically flexible, but the application scenarios are relatively limited.

    \item \textit{Firefox Extension\cite{firefoxExtension}}:

        It can be executed directly in the user's local browser, and access is controlled through the browser's permission management. Users must confirm permissions before installation and can adjust permissions after installation. The application supports direct modification or extension of browser functions to manipulate web content. Since developers upload the complete application code, a wide variety of extensions are available; but no online updates are provided.

    \item \textit{Chrome Extension\cite{chromeExtension}}:
   
     It executes in the user's local browser and permissions management is similar to Firefox. The user confirms the permissions before installation and after installation the user can view and adjust the permissions. chrome has thousands of plugins available and like Firefox, the developer uploads the complete application code and does not support online updates.

    \item \textit{ChatGPT Plugin}:

      Chatgpt executes remotely on a third-party server, accessing external services via API calls that can provide real-time data support during a session. Users can enable specific plugins during a session, with permissions limited to in-session access. There are multiple plugins available and the interaction process is influenced by the underlying Large Language Model (LLM).
  
\end{itemize}

\subsubsection{The Breadth and Flexibility of the ChatGPT Plugin Store}

In these comparisons, it is evident that the ChatGPT plugin system, as a new type of plugin platform, possesses three unique features:

\begin{enumerate}
    \item \textbf{Reliance on the Powerful LLM}: The integration of a large language model greatly extends the flexibility and functionality of plugins.

    \item \textbf{Wide Range of Functions and Permission Requirements}: ChatGPT plugins support diverse functions, covering multiple needs with varied permission requirements.

    \item \textbf{Real-Time API Interactions}: ChatGPT adopts a flexible online API interaction model, allowing users to call APIs in real-time during conversations, enabling diverse and dynamic interactions. This also allows relatively flexible plugin updates.
\end{enumerate}

While this high flexibility and real-time API calling make ChatGPT plugins outstanding in multiple scenarios, it also introduces complex security and privacy risks. Compared to traditional plugins, ChatGPT plugins require stricter management policies and security protocols to protect user privacy and data. Additionally, the potential impact of the LLM on plugin interactions and security must be carefully assessed.

\begin{table}[t]
\centering
\renewcommand{\arraystretch}{1.5} % 增加行高
\caption{Comparison of Plugin Systems across Multiple Dimensions}
\small
\begin{tabular}{|p{2.5cm}|p{3cm}|p{3cm}|p{3cm}|p{3cm}|}
\hline
 & \textbf{Notion API Integration} & \textbf{ChatGPT Plugin} & \textbf{Firefox Extension} & \textbf{Chrome Extension} \\ \hline
\textbf{Code Execution Location} & Third-party server, online remote execution & Third-party server, online remote execution & User's local browser, direct execution & User's local browser, direct execution \\ \hline
\textbf{Permission Management} & User grants access to specific pages or databases \newline OAuth 2.0 authorization & Real-time plugin activation by user \newline Session-limited access & Permissions confirmed before installation; adjustable post-installation & Permissions confirmed before installation; viewable and adjustable post-installation \\ \hline
\textbf{Integration \& Operation} & Data interaction via API \newline Data processed remotely and returned & API calls external services \newline Provides data in conversation & Directly modifies or extends browser functionality \newline Controls web content & Directly modifies or extends browser functionality \newline Controls web content \\ \hline
\textbf{Plugin Scope} & Notion application and database only & Dynamic range of services and functionalities in conversations & Browser and web functionality & Browser and web functionality \\ \hline
\textbf{Plugin Quantity} & Limited, mainly focused on specific applications & Moderate, diverse types & Vast, tens of thousands of extensions & Vast, tens of thousands of extensions \\ \hline
\textbf{Plugin Coverage} & Data management, cloud collaboration & Information processing, intelligent interaction, API integration & Browser experience optimization, ad blocking, developer tools & Browser experience optimization, ad blocking, developer tools \\ \hline
\end{tabular}
\end{table}
\clearpage

% +++++++++++++++++++++++++++++++++++++++++++++++++++++++++++++++++++++
\section{Chatgpt Plugin Architecture and Key Components}

% 在上一节中，我们探讨了Current State of Security in Plugin Systems，本节将专注于 ChatGPT 插件系统本身，深入分析其架构与运作流程。主要内容包括：

% 插件的工作流程：探讨平台与插件服务器间的交互方式，帮助理解用户请求如何通过安全的调用机制处理。

% 清单文件的关键字段：分析清单文件中的核心字段，如插件基本信息和权限设置，确保插件合规和可控。

% 插件 API 文件的结构：剖析 API 文件的组成，包括接口路径和数据响应模式，以展示插件和平台之间的标准化接口。

% 平台对插件商店的管理：概述 OpenAI 的插件管理策略，确保插件的安全性和合规性。

% 我们希望，通过分析 ChatGPT 插件系统的这些要素，帮助理解其在平台中的集成方式和安全保障。
In the previous section we looked at the Current State of Security in Plugin Systems, this section will focus on the ChatGPT plugin system itself, analysing its architecture and workflow in depth. Key points include:

\textbf{The Workflow of Chatgpt plugins:} exploring how the platform interacts with the plugin server to help understand how user requests are handled through the secure invocation mechanism.

\textbf{Structure in Manifest Files:} Analysing the core fields in the manifest files, such as the basic information and permission settings of the plugin, to ensure that the plugin is compliant and controllable.

\textbf{Structure of plugin API files:} Dissect the composition of API files, including interface paths and data response patterns, to demonstrate standardised interfaces between plugins and the platform.

\textbf{Platform management of the plugin store:} outlining OpenAI's plugin management strategy to ensure plugin security and compliance.

We hope that analysing these elements of the ChatGPT plugin system will help understand how it is integrated and secured within the platform.

\subsection{The Workflow of Chatgpt plugins}

In this process, the user sends a request to the platform, usually a text or an image. The platform parses the "description\_for\_model" field in the user's plugin repository and determines which specific plugin is being used and sends the request and specifics to the server where the plugin resides. The plugin runs on its server and returns the results to the CHATGPT platform. There is no direct interaction between the user and the plugin server in this process; it is done through the platform.

\begin{figure}[H]
    \centering
    \includegraphics[width=0.7\linewidth]{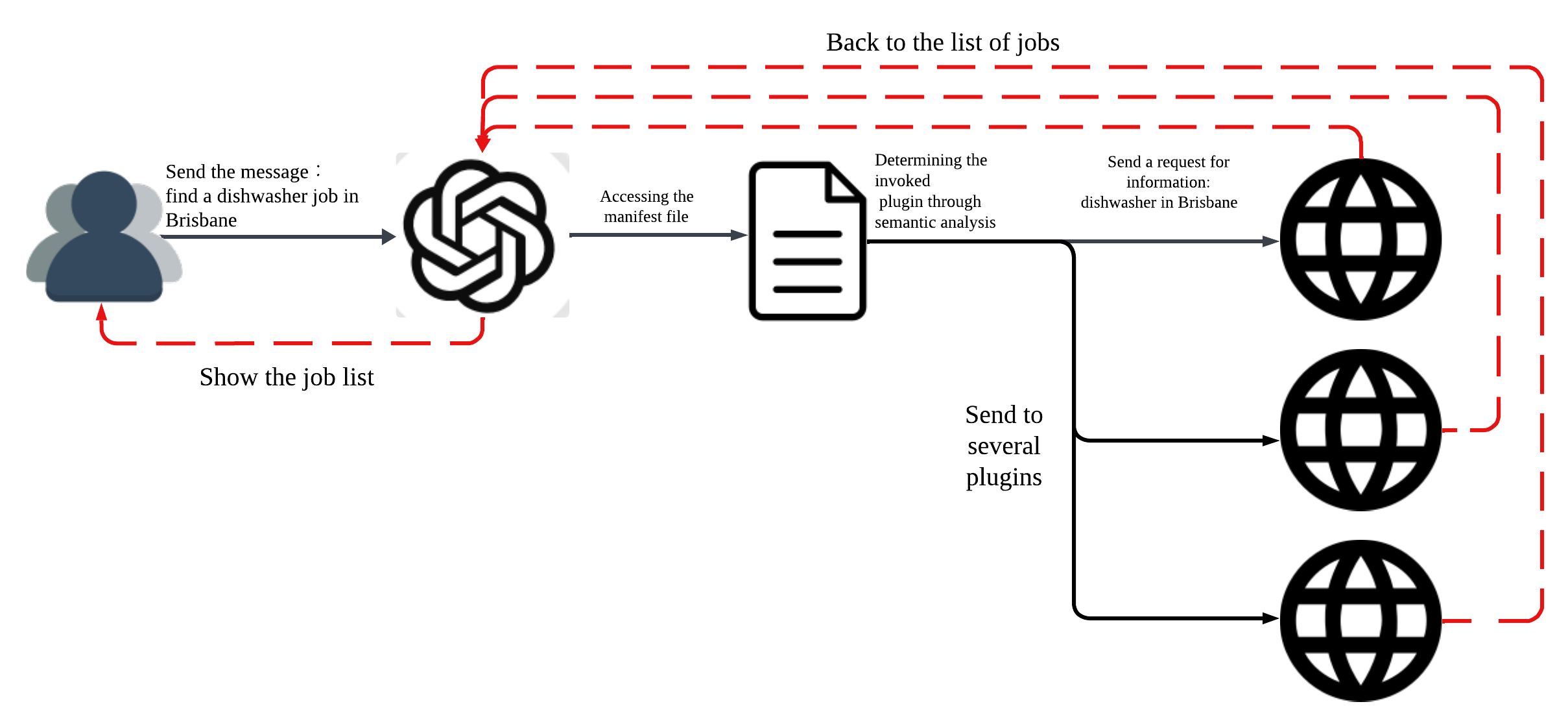}
    \caption{Workflow of ChatGPT plugin}
    \label{fig:enter-label}
\end{figure}

\subsection{Key Fields and Structure in Manifest Files}

% 在这一部分中，我们将讨论 JSON 结构和其内部的字段，这些字段包含了有效的可探查信息，以供我们进一步探讨 ChatGPT 插件的功能及其安全性。

% name_for_human：
% 这个字段用于展示给人类用户的插件名称，通常是一个简单易懂的名字。它帮助用户快速了解插件的功能或用途，特别是在插件商店或插件列表中进行浏览时。

% name_for_model:
% 这是提供给机器学习模型使用的名称。通常比 name_for_human 更简洁，以便于模型识别和处理该插件。在模型进行调用时，插件通过这个字段来识别

% auth：

% authorization_type，该字段用于定义插件的身份验证机制。不同的 authorization_type 会对插件的安全性、用户访问控制和与系统交互的方式产生显著影响，在chatgpt 插件中， authorization_type主要分为三个 类型：
    % None： 这种方式表示不需要身份验证，任何用户或系统都可以直接调用 API。由于不需要身份验证，用户可以直接访问资源，操作简便，但也因此存在安全隐患。
    
    % Bearer Token（持有者令牌）： Bearer Token 是一种常用的身份验证方式，授权方通过一个 token 来验证用户的身份。任何拥有该 token 的实体都可以访问相关资源。其主要优势在于灵活性高，可以在无需复杂身份验证的情况下快速授权访问，但也因此存在一定的安全风险，特别是在 token 泄露的情况下。
    
    % OAuth： OAuth 是另一种常见的身份验证方式，它允许用户通过第三方服务（如 Google 或 Facebook）登录。这种方式提供了更高的安全性，因为它不直接使用用户的登录凭据，而是通过授权服务器获取访问令牌。OAuth 能有效地保护用户信息，并减少因泄露而导致的风险。
    
% "scope":通常 OAuth 允许开发者通过 scope 指定访问范围

%"verification_tokens":指定了授权请求的内容类型，

% api：

% 该字段包含关于插件 API 的详细信息，常见的子字段包括：
% type：定义 API 的类型，常见的值为 "openapi" ，表示 API 是遵循 OpenAPI 规范
% url：插件 API 的文档 URL，通常提供了 API 的完整说明，包括可用的端点、请求和响应格式等。开发者可以使用此文档了解如何与该 API 进行交互。
% is_user_authenticated：一个布尔值，表示调用 API 是否需要用户认证。如果为 true，则用户在使用插件功能时必须经过身份验证。

% legal_info_url：

% 提供插件的法律信息链接，通常指向服务条款（Terms of Service, TOS）或隐私政策（Privacy Policy）等文档。这些文档确保用户在使用插件时了解相关法律责任和数据隐私条款。

%我们的讨论将集中在api信息相关(在第二层涉及)，包括API可触达，token权限对于泄露严重性的影响，而在第三层我们将集中讨论法律信息等一致性问题。

\begin{figure}[H]
\centering
\includegraphics[width=1\linewidth]{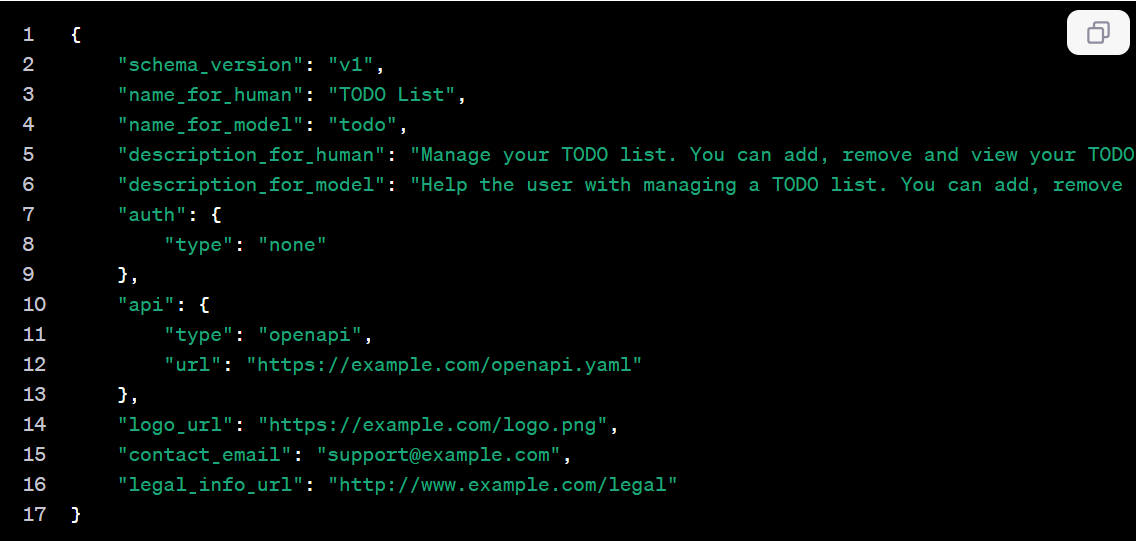}
\caption{\label{fig:manifest}Manifest Sample in Openai\cite{plugingetstart}}
\end{figure}

According to the official OPENAI documentation, each plugin should contain a json file hosted on the API's domain to connect to the plugin platform. The plugin platform looks for this file under the path "/.well-known/ai-plugin.json" or "/.well-known". This JSON file has a set of well-defined minimum fields\cite{plugingetstart}\cite{pluginsauthentication}.

In this manifest file, the plugin developer is asked to provide at least the name of the plugin, an associated description, API information, and legal information in order to standardize and enhance its readability, specifically\cite{plugingetstart}\cite{HossenMohammadSahinur2023MDiW}:

\subsubsection{name\_for\_human:} 
This field is used to present the name of the plugin to human users, usually a simple and understandable name. It helps the user quickly understand the function or purpose of the plugin, especially when browsing in the plugin store or plugin list.

\subsubsection{name\_for\_model:} 
This is the name provided for use with machine learning models. It is usually more concise than name\_for\_human to make it easier for the model to recognize and process the plugin. When the model makes a call, the plugin is recognized by this field.

\subsubsection{auth:} 
\paragraph{authorisation\_type:} Defines the plugin’s authentication mechanism. Different authorisation\_types can significantly impact the plugin's security, user access control, and system interaction. In ChatGPT plugins, there are three main types of \textit{authorisation\_type}:

\begin{itemize}
    \item \textit{None}: No authentication is required, allowing any user or system to call the API directly. While this simplifies access, it also presents a security risk due to the lack of access control.
    
    \item \textit{Bearer Token}: A common authentication method where the authorised party authenticates the user via a token. Any entity holding the token can access resources. The main advantage is flexibility, enabling quick access without complex authentication, but it can pose security risks if the token is compromised.

        \begin{figure}[H]
            \centering
            \includegraphics[width=1\linewidth]{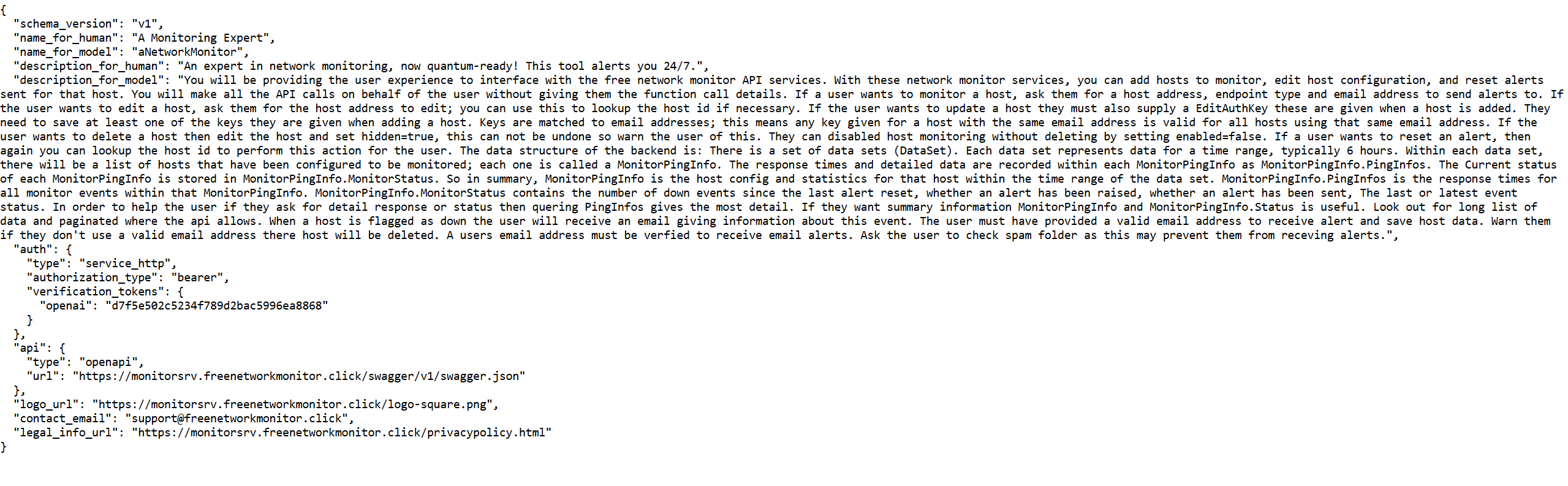}
            \caption{Bearer token example}
            \label{fig:bearer-token-example}
        \end{figure}
        
        \begin{figure}[H]
            \centering
            \includegraphics[width=1\linewidth]{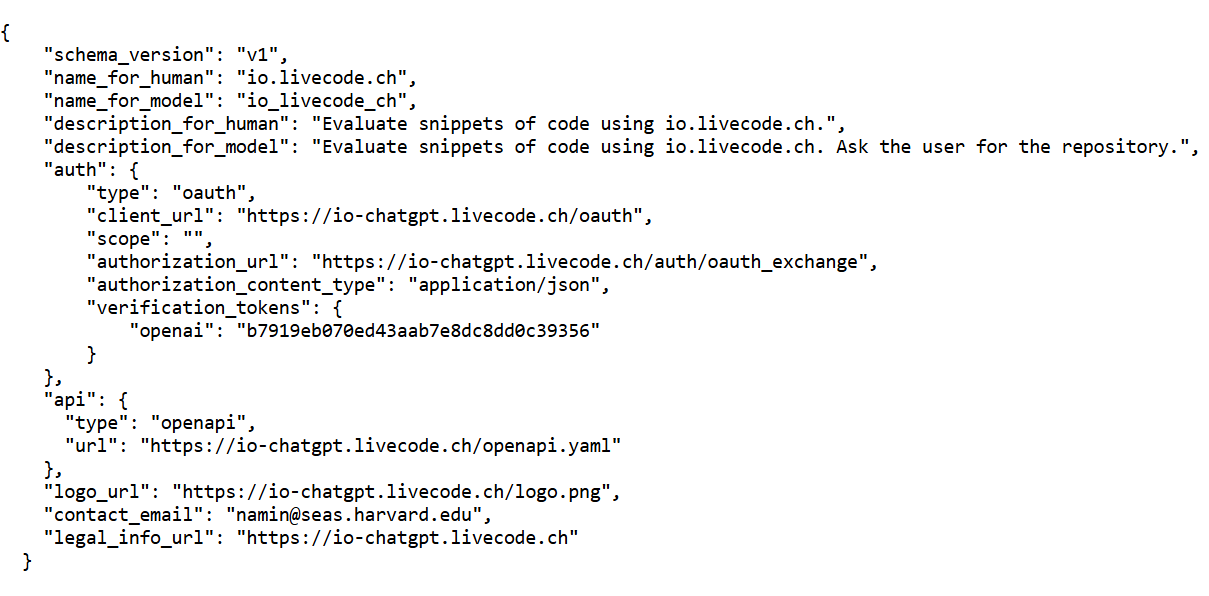}
            \caption{OAuth token example}
            \label{fig:oauth-token-example}
        \end{figure}

    \item \textit{OAuth}: Another widely used authentication method, allowing users to sign in through third-party services such as Google or Facebook. OAuth provides a higher level of security by not directly using the user's login credentials, instead obtaining an access token from an authorisation server, effectively protecting user information and reducing leakage risks.
\end{itemize}

\paragraph{scope:} Typically, OAuth allows the developer to specify access limitations through a defined scope. 

\paragraph{verification\_tokens:} Specifies the content type of the authorisation request. 

\subsubsection{api:} 
This field contains detailed information about the plugin API, with common subfields including:

\begin{itemize}
    \item \textit{type:} Defines the type of the API, typically set as "openapi", indicating that the API follows the OpenAPI specification.
    
    \item \textit{url:} The documentation URL for the plugin API, usually providing a comprehensive API description, including endpoints, request and response formats, etc. This URL enables developers to understand how to interact with the plugin API.
    
    \item \textit{is\_user\_authenticated:} A boolean indicating if user authentication is required for API access. If true, the user must be authenticated to use the plugin's functionality.
\end{itemize}

\subsubsection{legal\_info\_url:} 
Provides a link to the plugin's legal information, usually pointing to documents such as the Terms of Service (TOS) or Privacy Policy. These documents ensure that users are aware of their legal responsibilities and data privacy terms when using the plugin.

% Our discussion will focus on API-related information (covered in the second tier), including API reachability and the impact of token permissions on potential leaks. In the third tier, we will concentrate on consistency issues, such as the legal information provided.

\subsection{The Structure of Plugin's API File }
% 插件的 API 文件遵循 OpenAPI 规范，用于定义插件的基本信息、服务器路径、API 路径与操作、以及响应数据结构等内容。它通常嵌入到清单文件中，构成插件在平台中的接口说明。API 文件详细描述了插件的具体功能和数据交互方式，以确保插件能正确地集成到平台并提供所需服务。其基本结构如下：

% API 文件的顶层结构:包括 openapi、info、servers、paths 和 components。其中，openapi 字段指定 OpenAPI 版本，以确保文件格式的标准化；info 字段提供插件的基本信息，如名称、描述和版本号，帮助用户和平台理解插件的功能与用途；servers 字段定义插件服务器的 URL，作为 API 请求的基础路径；paths 字段描述插件的 API 路径和操作，是插件功能的核心部分；components 字段定义响应数据模式，确保返回的数据符合预期格式。

% 在 info 字段中，列出了插件的基本信息，如 title（插件名称）、description（插件的简要说明）和 version（插件的版本号），这些信息有助于用户和平台理解插件的功能和用途。servers 字段通过 url 指定插件服务器的位置，为 API 请求提供目标地址。

% API 路径与方法（paths） 部分定义了每个路径对应的具体功能操作，通常包含以下 HTTP 请求方法：GET、POST、PUT 和 DELETE，分别用于执行数据的检索、创建、更新和删除操作。数据响应结构（components） 部分则定义了 API 的响应数据模式，以确保返回的数据结构规范且一致。components 中包含多个 schemas，用于描述不同类型的数据格式。

% 这种结构使得插件能够在平台上实现标准化和一致性的交互，有助于提高插件的可用性和数据传输的安全性。

The API file for a plugin follows the OpenAI specification and defines basic information about the plugin, server paths, API paths and operations, and response data structures. It is usually embedded in a manifest file and constitutes the plugin's interface description in the platform.The API file describes in detail the plugin's specific functionality and data interactions to ensure that the plugin integrates correctly into the platform and provides the required services. The basic structure is as follows:\\

\begin{figure}[H]
\centering
\includegraphics[width=1\linewidth]{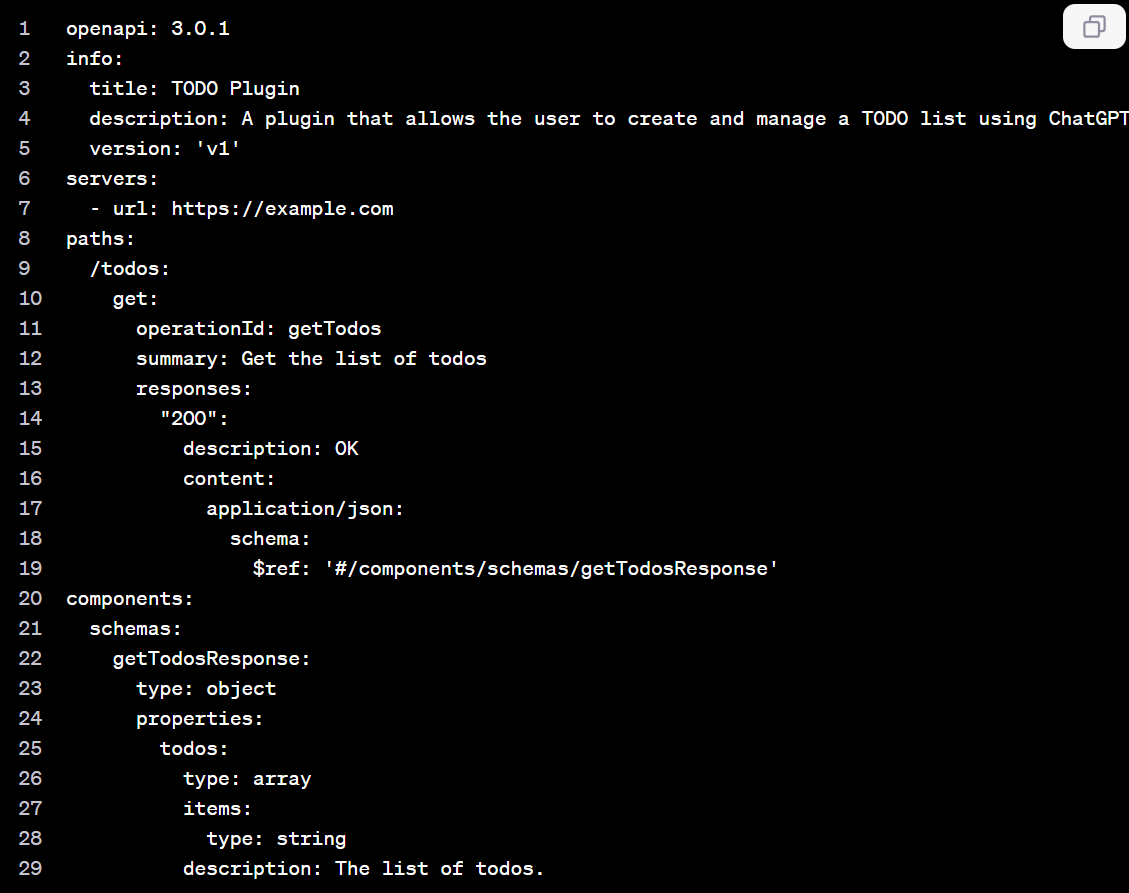}
\caption{\label{fig:manifest}Openapi Sample\cite{plugingetstart}}
\end{figure}

\subsubsection{The top-level structure of the API file:} openapi, info, servers, paths and components, where the openapi field specifies the OpenAPI version to ensure the standardisation of the file format, the info field provides basic information about the plugin, such as the name, description, and version number, to help the user and the platform to understand the plugin's functionality and usage, and the servers field defines the plugin's functionality and usage.

% \subsubsection{The info field:} lists basic information about the plugin, such as title (plugin name), description (brief description of the plugin), and version (version number of the plugin), which helps the user and the platform to understand the plugin's functionality and usage. servers field specifies the location of the plugin's server via the url, which provides a destination for the API request. address.

\subsubsection{The API paths and methods section:} defines the specific functionality of each path, which typically includes the following HTTP request methods: GET, POST, PUT, and DELETE, which are used to retrieve, create, update, and delete data, respectively.

\subsubsection{The components section:} defines the API's response data schema to ensure that the returned data is structured in a standardised and consistent manner. components contains multiple schemas that describe different types of data formats.

This structure allows plugins to have standardised and consistent interactions across the platform, which helps to improve the usability of the plugin and the security of data transfers.

\subsection{Openai's Management of the Plugin Store}
In the platform's official regulations, in addition to the basic content-based requirements such as content policy and brand guidelines that must be followed\cite{reviewprocess}, the plugin store also sets some specific requirements: For example, not interfering with search rankings through variant beginnings, and beginning with the meaningless quantifier 'A' to make the plugin rank in the front of the test\cite{reviewprocess}; for plugins executed worldwide, the Oauth authentication is required, etc., in order to protect the security of user identity and data confidentiality\cite{reviewprocess}\cite{chatplugin}.

Meanwhile, the platform requires developers to safeguard the security of API plugins and requests against possible vulnerability behaviors; This includes ensuring that the use of plugins does not bring security vulnerabilities, threats, or malware infringements to third parties, which requires developers to take appropriate security measures when developing and maintaining plugins to protect users and the entire platform from security threats\cite{PluginsandActionsTermss}.

According to the Openapi specification, plugins can choose specific endpoints to present their functionality to users, follow the traditional Openapi format, and set up simple public API local proxies\cite{pluginsauthentication}.

\section{Current Problems and Gaps}
Based on the existing research, we will further explore the security issues that may occur in the ChatGPT plugin store and analyze their impact and causes. Although the ChatGPT plugin platform is designed to be a plugin store that is generative, flexible, and extensive, there are still the following issues that have not been explored in depth in actual applications:

\subsection{Information Exposure Issues}
Public visibility of Manifest files: We found that a significant portion of manifest files are placed in public web domains without any protection measures, which not only exposes API-related information to the public domain, but also leaks specific documents as a result. This situation directly increases the security risk and provides clues for potential attackers to find attack targets.
\subsection{Permission Granting}
OpenAI allows plugins to log in users through redirection\cite{pluginsauthentication}, a process in which the user often grants the plugin too many permissions without realizing it. The covert nature of this authorization process makes it difficult for users to notice what permissions they have granted, leading to the possibility that they may have inadvertently given third-party plugins too much access to their data, which significantly increases the risk to user privacy and data security.
\subsection{Other Unusual Competitive Practices} 
Although OpenAI explicitly prohibits the use of unfair competition tactics, such as adding meaningless quantifiers to the name of a plugin in order to gain a more prominent position in the search result\cite{PluginsandActionsTermss}, this type of behavior has not been completely curbed. Such tactics not only distort users' search results, but may also lead users to download and use unsafe or low-quality plugins
\subsection{Detection of Isolation}
Although the CHATGPT platform theoretically prevents information leakage and security issues through isolation mechanisms, in practice the mechanisms for detecting and enforcing plugin isolation do not seem to be sufficient. Effective isolation mechanism should be able to ensure that the plugin execution will not access or affect the user's other data and plugin operating environment, at present, we initially found that there are bypassing the platform can be directly interacted with the plugin through the API.

\section{Motivation}

% 我们将专注于以下研究脉络：

% 三层一致性与安全性验证框架：构建并应用一个三层框架，系统性地验证 ChatGPT 插件系统中的一致性和安全性，以识别和评估其潜在的风险和漏洞。

% 两种情境下的案例研究：通过分析两个实际使用情境中的插件情况，详细探讨系统在不同应用场景中的安全表现和风险点。

% 七个泄露与一致性问题：重点分析插件系统中七个关键的泄露和一致性问题，包括清单文件的公共可见性、API 权限管理不当、认证机制不一致、无保护的 API 端点、授权令牌管理缺失、法律信息不全、以及插件隔离机制的不足。

% 两个 ChatGPT 演变时期：对比分析 ChatGPT 插件系统在两个不同时期的安全策略和管理模式，以探讨其安全性改进的进展及未来优化方向。

As recounted in the introduction, current research focuses on the security of LLMs and pays insufficient attention to the problems of the Plugin system itself. The aim of this study is to explore the security vulnerabilities of the CHATGPT Plugin system and to propose improvements in the hope of improving the system's security.

Identify and describe: the major security vulnerabilities and improper exposures in the CHATGPT Plugin system.

Assess: The impact of these vulnerabilities on users and overall system security.

Propose and test improvements: aimed at enhancing the security of the CHATGPT plugin system and reducing security threats.\\

% Based on this, We will focus on the following research vein:

% \textbf{Three-layer consistency and security validation framework:} Construct and apply a three-layer framework to systematically validate the consistency and security in the ChatGPT plugin system to identify and assess its potential risks and vulnerabilities.

% \textbf{Two Case studies:} By analysing the plugin in two real-life usage scenarios, we explore in detail the security performance and risk points of the system in different application scenarios.

% \textbf{Seven Leakage and Consistency Issues:} Focuses on analysing seven key leakage and consistency issues in the plugin system, including public visibility of manifest files, improper management of API permissions, inconsistent authentication mechanisms, unprotected API endpoints, missing authorisation token management, incomplete legal information, and deficiencies in plugin isolation mechanisms.

% \textbf{Two ChatGPT evolution periods:} Comparative analysis of the security policies and management models of the ChatGPT plugin system in two different periods to explore the progress of its security improvement and future optimisation direction.

% \clearpage
%增加不同商店的比较,要回去盘下逻辑，有的，后面的，可能要放到这里

\chapter{Methodology}
In this section, we create a multi-class approach to identify security issues in the Chatgpt plugin system, our focus is on the reasonableness of Oauth request permissions, the exposure of manifest files and the security risks of api files.We will make use of python's selenium library and scripts to analyse and collect data, evaluate security configurations and monitor security vulnerabilities.

\section{Consistency and Security Verification Framework}
We will test the security of the plugin using a multi-layered API security testing framework designed to identify and mitigate potential security threats and ensure data integrity and security of the API. We present three main testing layers: 

\begin{itemize}
    \item {Manifest Analysis Layer}  
    \item {API Request Analysis Layer}      
    \item {Consistency and Integrity Analysis Layer}  
\end{itemize}

Each of which provides specialized testing strategies and methods for different security aspects of the API.

\begin{figure}[H]
    \centering
    \includegraphics[width=0.77\linewidth]{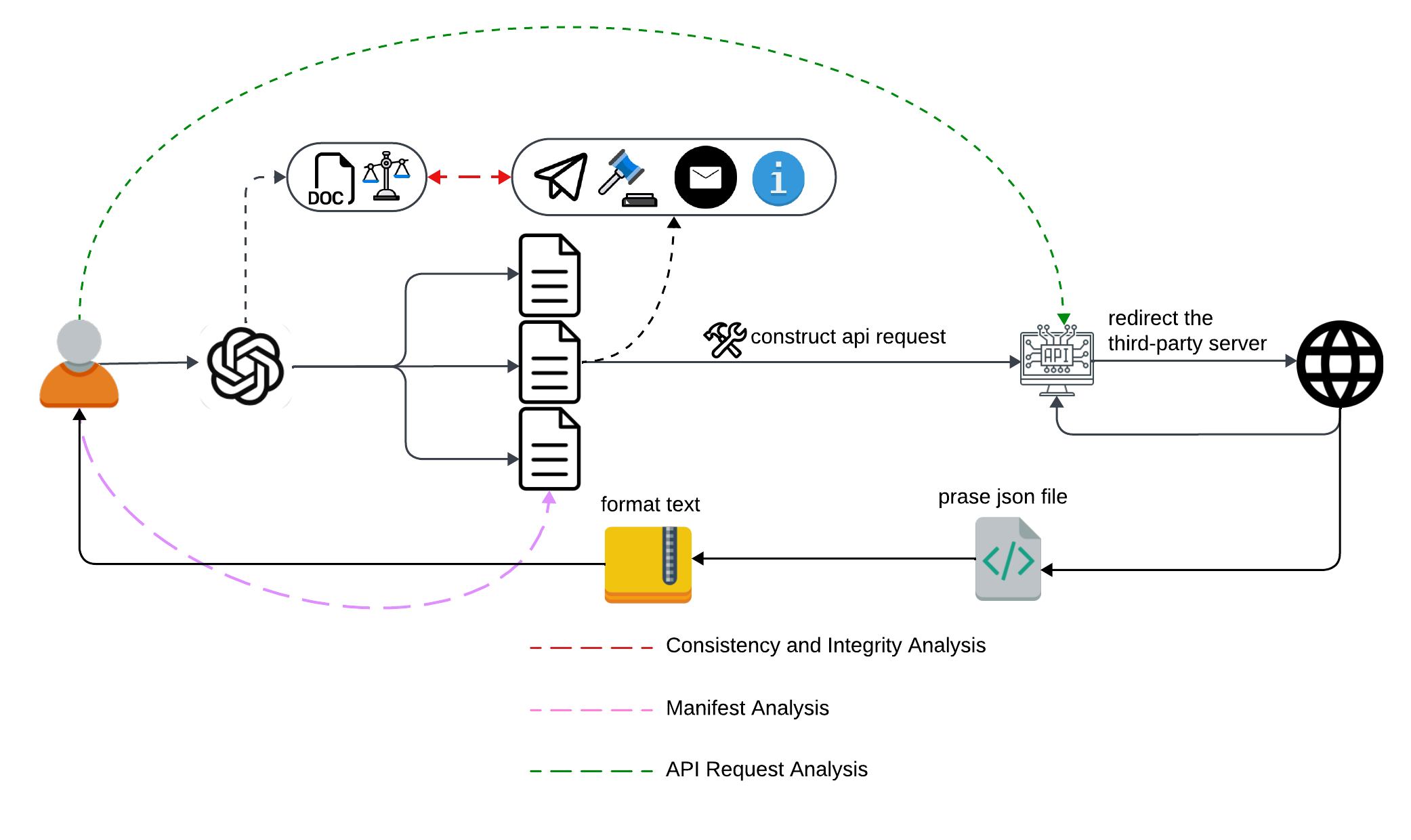}
    \caption{Security Checking Model}
    \label{fig:enter-label}
\end{figure}
\subsection{Manifest Analysis Layer}
\subsubsection{The theory of Manifest Analysis Layer}

Manifest file exposure, this dimension we mainly aim to is to identify potential information leakage, especially the leakage of configuration file manifest file to determine whether the plugin is protected or not. And whether the plugin is protected against unauthorised malicious access.

We will use Python package\cite{beautifulsoup2020}related automation tools to scan the subpages of the relevant domains. According to the official Chatgpt documentation, the manifest file of the API needs to be placed in the corresponding URL for the Chatgpt platform to identify the access. Unauthorised access attempts or public access by non-platforms should be identified and denied.

\subsubsection{Mathematical Formulation}
Let $S$ be the set of all files, $E$ the subset of exposed files identified by automated tools, and $P$ the subset of protected files not in $E$. Let $L$ denote the storage location, which could be an external site or network address.

The exposure detection can be represented as:
\begin{equation*}
    E = \{s \in S \mid \text{access to } s \text{ is unauthorized}\}
\end{equation*}

The non-exposure analysis is represented as:
\begin{equation*}
    P = S \setminus E
\end{equation*}

Verification of storage location for each protected file:
\begin{equation*}
    \forall s \in P, \exists l \in L \text{ such that } l \text{ is secure}
\end{equation*}

% 写怎么做的，步骤和探查路径
\subsubsection{Steps for Manifest Analysis Layer}
\begin{figure}
    \centering
    \includegraphics[width=1\linewidth]{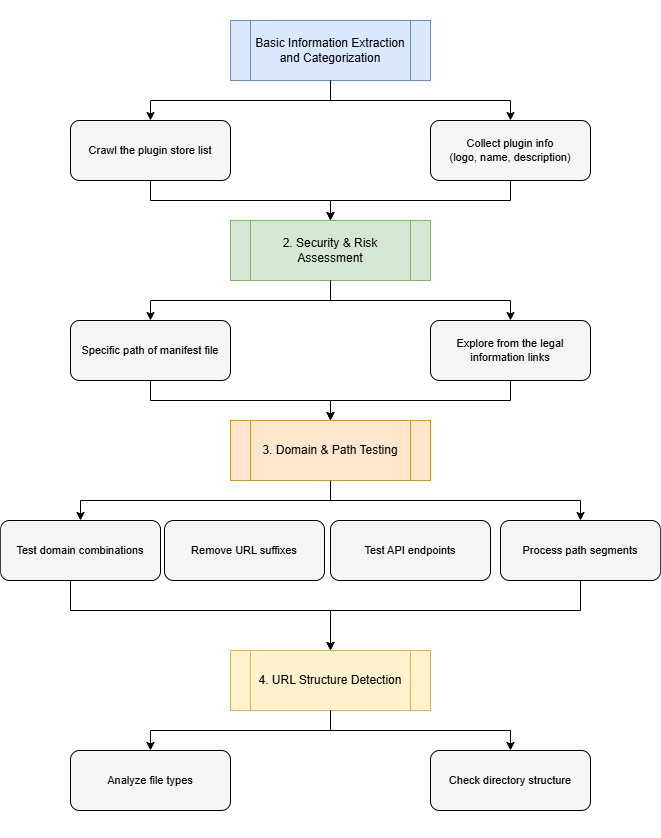}
    \caption{Steps for detecting leaks of manifest file}
    \label{fig:enter-label}
\end{figure}

\paragraph{Basic Information Extraction and Categorization of ChatGPT Plugins}
% 1. 基础插件信息的爬取与分类
% 为了全面了解 ChatGPT 插件的生态系统，我们首先着手爬取所有插件的基础信息。爬取的基础信息包括插件的 logo、插件名称（title）、面向用户的名称和法律信息链接。这些消息出现在插件商店主页，是我们主要的信息获取来源。
 In order to get a full picture of the ChatGPT plugin ecosystem, we first set out to crawl the base information of all plugins. The basic information crawled includes the plugin's logo, plugin name (title), user-facing name, and links to legal information. These messages appear on the plugin store homepage and are our primary source of information acquisition.
 
Finally, we collected 1033 chatgpt plugins.

\paragraph{Information Leakage Risk Assessment via Manifest File Analysis}

% 在收集了足够多的插件基本信息后，我们正式开始了插件安全和信息泄露风险评估。我们主要通过抓取和分析 ChatGPT 插件的清单文件（即 manifest 文件）来推测插件在信息交互中的泄露风险。

% 根据 OpenAI 的规定，每个插件的清单文件应存储在 / .well-known/ai-plugin.json 或 / .well-known 路径下，以便 ChatGPT 能正确识别和调用插件。因此，我们尝试了不同的域名和路径组合，以找到可能的清单文件。
% 我们将从法律信息链接为出发点，试图探究manifest文件的泄露
 After collecting enough basic information about the plugin, we formally started the plugin security and information leakage risk assessment. We mainly crawl and analyse the manifest files of ChatGPT plugins to speculate the leakage risk of plugins in information interaction.

According to OpenAI, the manifest file of each plugin should be stored under the path / .well-known/ai-plugin.json or / .well-known, so that ChatGPT can correctly identify and invoke the plugin. Therefore, we tried different combinations of domains and paths to find possible manifest files.

We will try to explore the manifest file leaks from the legal information link as a starting point

\paragraph{Testing Domain and Path Combinations for Manifest Discovery}
% 3. 域名和路径组合的测试
% 我们通过编写 Python 脚本，对不同域名和子域名后缀进行组合，假设它们可能是 manifest 文件的地址。例如，域名的拆解、重新组合、移除特定子目录后缀等，都是我们用来探索这些文件的常用方法。我们还通过逐步分割网址路径，尝试提取前三层或前两层目录，并将其与子路径进行拼接，以期能够找到潜在的 manifest 文件。
% 静态 API 目录探测：
% 为了提高识别潜在应用程序接口目录的准确性，我们首先去除了 URL 后缀，将探测重点放在域名和预定义路径后缀组合上。这种方法旨在减少不必要后缀造成的干扰，优化静态资源目录的位置。

% 域名和预定义路径后缀的多种组合：
% 根据域名，我们提取路径的前两段或前三段，并进行系统测试，以确定可能的清单文件存储路径或其他关键资源。这种路径分段可逐步缩小探索范围，从而提高文件检索的准确性。

% 文件类型和 URL 结构检测：
% 在文件类型检测中，我们重点剔除常见的文件类型后缀，如.php、.txt、.htm、.pdf 等，或直接检测其对应的原始文件。同时，/pages、/us、/en、/static 等目录结构也在我们的检测范围内，通过去除特定的目录后缀，进一步扩大可能的路径组合。这种方法不仅提高了检测覆盖率，还有效避免了路径后缀产生的障碍。
%这种方法取得了部分成功，相当一部分插件被发现存在清单文件泄露情况，同时，有些插件由于各种原因，无法寻找有效的网址，我们将一一分析这些情况

Based on links to legal provisions, We do this by writing Python scripts that combine different domain names and subdomain suffixes, assuming that they may be the addresses of manifest files. For example, disassembling domain names, recombining them, and removing specific subdirectory suffixes are common methods we use to explore these files. We also attempted to extract the first three or two levels of directories by progressively splitting the URL paths and splicing them with subpaths in an attempt to be able to find potential manifest files.

\subparagraph{Static API Directory Detection:}
In order to improve the accuracy of identifying potential API directories, we first remove URL suffixes and focus our probing on combinations of domain names and predefined path suffixes. This approach is intended to reduce interference caused by unnecessary suffixes and optimise the location of static resource directories.

\subparagraph{Multiple combinations of domain names and predefined path suffixes:}
Depending on the domain name, we extract the first two or three segments of the path and perform systematic tests to identify possible inventory file storage paths or other critical resources. This path segmentation allows for progressively narrower exploration, thus improving the accuracy of file retrieval.

\subparagraph{File type and URL structure detection:}
In file type detection, we focus on eliminating common file type suffixes, such as .php, .txt, .htm, .pdf, etc., or directly detecting their corresponding original files. At the same time, directory structures such as /pages, /us, /en, /static, etc. are also in our detection scope, further expanding the possible path combinations by removing specific directory suffixes. This approach not only improves the detection coverage, but also effectively avoids the obstacles created by path suffixes.

This approach has been partially successful, and a significant number of plugins have been found to have leaked manifest files, while some plugins have been unable to find valid URLs for a variety of reasons, which we will analyse one by one.

% ++++++++++++++++++++++++++++++++++++++++++++++++++++++++++++++++++++++++++++++++++++++++
\subsection{API Request Analysis Layer}

\subsubsection{Purpose}
In this dimension we focus on API security testing. We will obtain relevant api information from the manifest file, such as feasible request constructs and corresponding paths.
For plugins that have obtained the api information, we will simulate the chatgpt platform to send requests to test if the plugin has the ability to distinguish the source of requests. We will also verify how the API reacts when receiving unauthenticated requests, bypassing the chatgpt platform by sending requests to the plugin API via selenium\cite{selenium} disguised as a third-party human user.

Our testing will encompass the following levels:
\paragraph{Unique Identifier Verification:}We will use the inclusion of the bearer header to verify that the API endpoint's token authentication mechanism is working properly, where we will also have to take into account the identifier encryption that comes with the chatgpt plugin.
\paragraph{OAuth authorisation detection:} for plugins authorised via OAuth third-party redirection, we will simulate the entire redirection process to assess whether the API's OAuth authentication is working properly and whether there are any other security vulnerabilities.
\paragraph{API Information Testing:}We will directly detect the API request path without authorisation requirements, assess his importance in the whole API, and assess his risk level if it is related to the core function of user information leakage.

\paragraph{Simulated Request Testing:} We test the robustness and security response of the API by constructing and sending non-normal paths or malicious requests several times to simulate the behaviour of potential attackers.

% \paragraph{Response validation Test:} for the normal response to our request API, we will compare his response to the request of unknown sources and the response through the LLM processing, in order to determine the threat level of the API exposure, if the response is very similar, it can be judged that the API does have a problem of exposure.

\subsubsection{Evaluation Criteria}
Unauthenticated requests should not receive the same access or results as authenticated requests. In theory, all requests should not bypass the Chatgpt platform to get a response. Therefore, once a response is obtained, we need to assess the danger and potential vulnerability crisis it may pose.

\subsubsection{Mathematical Formulation}
Let $R$ represent all requests sent to the API, $R_{token}$ the subset of requests that include a token, and $R_{no\_token}$ the subset without a token. Let $E$ be the set of requests that are denied access.
\begin{align*}
    R_{token} &= \{ r \mid r \text{ includes a Token}\} \\
    R_{no\_token} &= \{ r \mid r \text{ does not include a Token}\} \\
    E &= \{ r \mid r \text{ is denied access}\} \\
    \forall r \in R, & \quad r \in E \text{ confirms that the system} \\
    & \quad \text{effectively denies all unauthorized requests}
\end{align*}

\subsection{Consistency and Integrity Analysis Layer}

\subsubsection{Purpose}
This part focuses on the acquisition of sensitive data through the consistency of the data acquired by the API and the internal database, in order to avoid security and compliance issues caused by data inconsistency. For the store's search strategy (ranking by first letter), some plugins use meaningless quantifiers or other terms in order to get a top ranking. For legal information, which must be disclosed to users, some developers choose to replace the legal information page with other information, thereby violating users' right to know.

\subsubsection{Method}
Information consistency tests: The data returned by the API is compared to the corresponding data in the database by writing and executing a series of automated scripts. These tests are designed to verify the completeness, accuracy and timeliness of the data.

\subsubsection{Mathematical Formulation}

First, define the match function \( \text{match} \), which directly checks if two data items are equal and returns a Boolean value:
\begin{equation*}
    \text{match}(d_{\text{api}}, d_{\text{db}}) = 
    \begin{cases}
        \text{true} & \text{if } d_{\text{api}} = d_{\text{db}} \\
        \text{false} & \text{otherwise}
    \end{cases}
\end{equation*}

% Define the set I containing all mismatched data pairs
Then, define the set \( \mathcal{I} \) containing all mismatched data pairs, where the data from the API does not match the data in the database, annotated with developer information:
\begin{align*}
    \mathcal{I} &= \left\{ (d_{\text{api}}, d_{\text{db}}) \mid d_{\text{api}} \in \mathcal{D}_{\text{api}}, d_{\text{db}} \in \mathcal{D}_{\text{db}}, \right. \\
    & \quad \left. \text{match}(d_{\text{api}}, d_{\text{db}}) = \text{false}, \right. \\
    & \quad \left. \text{dev}(d_{\text{api}}) = \text{dev}(d_{\text{db}}) \right\}
\end{align*}

% Count discrepancies per developer
Lastly, count the discrepancies per developer:
\begin{align*}
    \text{Discrepancies}_{\text{Developer}_{\text{id}}} &= \text{count}(\left\{ (d_{\text{api}}, d_{\text{db}}) \in \mathcal{I} \mid \right. \\
    & \quad \left. \text{dev}(d_{\text{api}}) = \text{Developer}_{\text{id}} \right\})
\end{align*}

\section{Web Crawling Techniques}
We follow previous approaches and adopt a series of automated techniques, including automatically crawling web resources, performing request redirection and modification, and simulating user interactions\cite{RenMengxia2023WAGC}. These techniques are combined into a comprehensive set of strategies to deeply analyze and evaluate the security performance of plugin systems.

We plan to automatically crawl by BeautifulSoup\cite{beautifulsoup2020}, identify and count the number of files in publicly accessible manifest files and API's through web crawler scripts. The purpose of this step is to assess the level of exposure of these files and the security of the sensitive information in them. We will modify the interaction method of the chatgpt platform, try to interact and determine whether there is an API that can bypass the ChatGPT platform and interact directly with the Plugin server to determine the closure and stability of the interaction..

\section{Data Analysis}
Based on the security checking model, we will use the above tools to organize and analyze the collected data, quantitatively analyze all frequency of file exposure and Oauth permissions, and combine it with qualitative analysis to evaluate potential security risks. All data analysis strictly follows privacy protection standards.

% \subsection{Designing Attack Scenarios}
% We have learned from the experience of our predecessors in using attack taxonomies and threat models to construct a persistent, updatable, and evolvable attack model.And  then we carefully categorizing the usage and characteristics of plugins, replacing the attack categories one by one to comprehensively examine their security\cite{IqbalUmar2023LPSA}.

% Based on the analysis of potential security threats to the ChatGPT plugin system, we plan to design a series of simulated attack scenarios including, but not limited to, privilege abuse, API leakage, data tampering, and denial-of-service attacks. These scenarios are designed to simulate attacks that may be launched by malicious users or automated scripts.

% We plan to use our own coded tools and scripts to execute the simulated attacks, while closely monitoring the response and behavior of the system. To minimize the impact on the actual runtime environment, we would like to perform these tests in a sandbox environment. Finally, the data collected during the simulated attacks are used to evaluate the security performance of the system, especially the protection against specific attack vectors. By analyzing the success of the attacks, we can identify the weaknesses of the system.
%%%manifest layer --6-10页
% -- 是什么（解释manifest字段）
% -- 怎么做的（步骤和路径探查的努力、）
% -- 得到的结果（成功，失败）
% -- 方法评估（路径探查的细化和容错，做出的努力，应对策略，缺点与局限性）---评估没写

%结果的分析和讨论
\chapter{Results and Discussion}

% 我们将整个研究工作划分为两个主要时期，第一个时期是在 4 月之前，这一阶段我们主要专注于对初代 ChatGPT 的插件商店及其插件功能的测试和观察。在这一阶段，由于 GPTs 尚未正式上市，ChatGPT 的插件呈现方式仍然依赖插件商店进行展示。插件商店内的插件以字母顺序排列，并通过分页进行展示。这个阶段的插件商店功能相对简单，仍处于早期开发阶段，因此我们能够更容易地对其进行爬取和分析。
We divided the whole research work into two main periods, the first one was before April, in which we mainly focused on testing and observing the plugin store and its plugin functionality of the first generation of ChatGPT. In this phase, the presentation of ChatGPT's plugins still relied on the plugin store for presentation, as GPTs were not yet officially available. Plugins in the plugin store are arranged in alphabetical order and displayed through pagination. The plugin store at this stage is relatively simple in its functionality and is still in an early stage of development, so we were able to crawl and analyse it more easily.

In our exploration of the three-layer framework, we identified the following eight security and consistency issues:

\begin{itemize}
    \item \textbf{Manifest File Leakage}: Manifest files are exposed on public URLs, resulting in the leakage of model, API, and legal information.

    \item \textbf{API Request Non-Token Interaction}: Information can be accessed without verification, leading to potential data leakage and unauthorized requests.

    \item \textbf{API Request Authorization Bypass}: Specific requests can be constructed to bypass authorization mechanisms.

    \item \textbf{Token Leakage}: Leaked tokens allow attackers to fully access the API, posing significant security risks.

    \item \textbf{Misleading Operation}: Discrepancies between the user-visible name and the model name may cause users to inadvertently authorize plugins incorrectly.

    \item \textbf{Malicious Spoofing}: The name field is sometimes misused to obscure the true intent of a plugin, misleading users into trusting unsafe plugins.

    \item \textbf{Privilege and model abuse}: Some plugins exploit name inconsistencies to increase click-through rates, which may lead users to grant excessive permissions.

    \item \textbf{Misleading Legal Information}: Certain plugins provide misleading or incomplete legal information, concealing important details from users.
\end{itemize}

In the following part, we will delve into each of these issues, providing detailed analysis. We will also incorporate findings from Case Studies 1 and 2 to assess the impact and prevalence of these security concerns.

\section{Results and Discussion in Manifest Layer}

% 在本节中，我们进行了多次重复试验，以确保结果的鲁棒性。为提升检测的准确性，我们还采用了多种方法对路径进行清洗，去除无效后缀，并尝试了不同的路径组合，以尽可能全面地评估各插件清单文件的可访问性和保护水平，最后，我们得到的结果如下
In this section, we performed several repetitive tests to ensure the robustness of the results. To improve the accuracy of the detection, we also used several methods to clean the paths, remove invalid suffixes.

Finally, we obtained the following results:

\begin{table}[h]
    \centering
    \caption{Accessibility of URLs}
    \begin{tabular}{@{}lcc@{}}
        \toprule
        \textbf{Category}   & \textbf{Accessible} & \textbf{Inaccessible} \\ \midrule
        success             & 373                 & -                     \\
        Hidden redirects    & 104                 & -                     \\
        OpenAI              &                     & 12                    \\
        Google doc          & -                   & 6                     \\
        Github              & -                   & 19                    \\
        Native URLs         & -                   & 518                   \\ \bottomrule
    \end{tabular}
\end{table}

%得到的结果（成功，失败）
\subsection{Results of Manifest File Accessibility Checks}
% 结论与数据总结
% 通过对 manifest 文件路径的穷尽式探索，我们最终找到了 373 个有效的 manifest 文件。这些 manifest 文件可以展示相关的 OpenAI 清单文档，为我们后续的安全性分析提供了重要的基础数据。

% 同时，我们也统计了无法访问的 manifest 文件相关数据：

% 非有效域名：我们发现共有 37 个非有效域名，包括 Google Doc 和 OpenAI GitHub 下的地址。其中，Google Doc 地址共有 19 个，占比 1.8%；OpenAI GitHub 地址共有 12 个，占比 1.1%；谷歌相关地址共有 6 个，占比 0.5%。

% Native URLs are unreachable：我们将那些即使经过多次路径组合尝试仍无法访问的原生网址归类为“原生网址不可达”。这些网址通常返回 403、404 或 406 等拒绝访问的错误代码。

% Hidden redirects：这些地址虽然可以通过路径组合访问，但由于返回的网页内容被其他信息覆盖，manifest 文件无法直接访问,我们发现了此类隐藏重定向网址104个,占比10.0%。

% 我们通过对这些数据的进一步分析，总结出了 manifest 文件在不同场景下的可达性情况，并对其中存在的安全性问题进行了详细讨论。这些结果为我们未来的工作提供了重要的参考依据。

\textbf{Accessible Files:}Through an exhaustive exploration of manifest file paths, we found 373 valid manifest files. These manifest files can display the relevant OpenAI manifest documents, which provide important basic data for our security analysis.\\
\textbf{Inaccessible Files:}We also counted the number of manifest files that were inaccessible:

\begin{itemize}
    \item \textbf{Non-valid domains:} We found 37 non-valid domains, including Google Doc and OpenAI GitHub addresses. Among them:
    \begin{itemize}
        \item 19 GitHub  addresses, accounting for 1.8\%.
        \item 12 Openai addresses, accounting for 1.1\%.
        \item 6 Google doc addresses, accounting for 0.5\%.
    \end{itemize}
    
    \item \textbf{Native URLs are unreachable:} We classify native URLs as ‘native URLs unreachable’ if they cannot be accessed even after multiple attempts to combine paths. These URLs typically return an access denied error code such as 403, 404, or 406.
    
    \item \textbf{Hidden redirects:} Although these addresses can be accessed through path combinations, the returned page content is covered by other information and the manifest file cannot be accessed directly. We found 104 such hidden redirects, accounting for 10.0\%.
\end{itemize}

We further analysed these data to summarise the accessibility of manifest files in different scenarios, and discussed the security issues in detail. These results provide important references for our next step work.
%（分析manifest文件的描述，和404 406这些的占比）

% 错误讨论
\subsection{Limitations Due to Missing or Invalid External Website Links}

\begin{figure}[H]
    \centering
    \begin{tabular}{cc}
        \begin{subfigure}{0.4\textwidth}
            \centering
            \includegraphics[width=\linewidth]{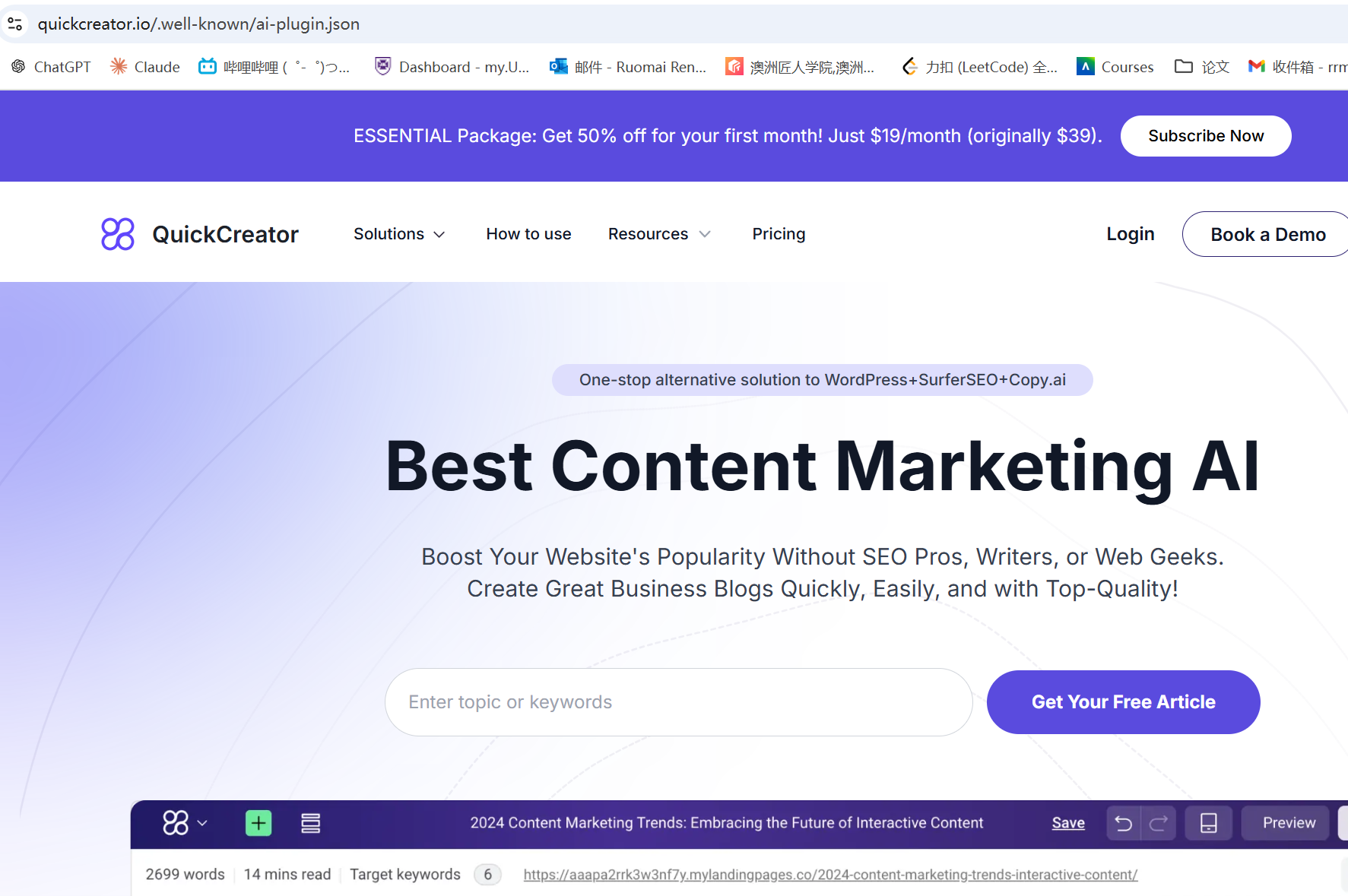}
            \caption{Hidden redirects}
        \end{subfigure} &
        \begin{subfigure}{0.4\textwidth}
            \centering
            \includegraphics[width=\linewidth]{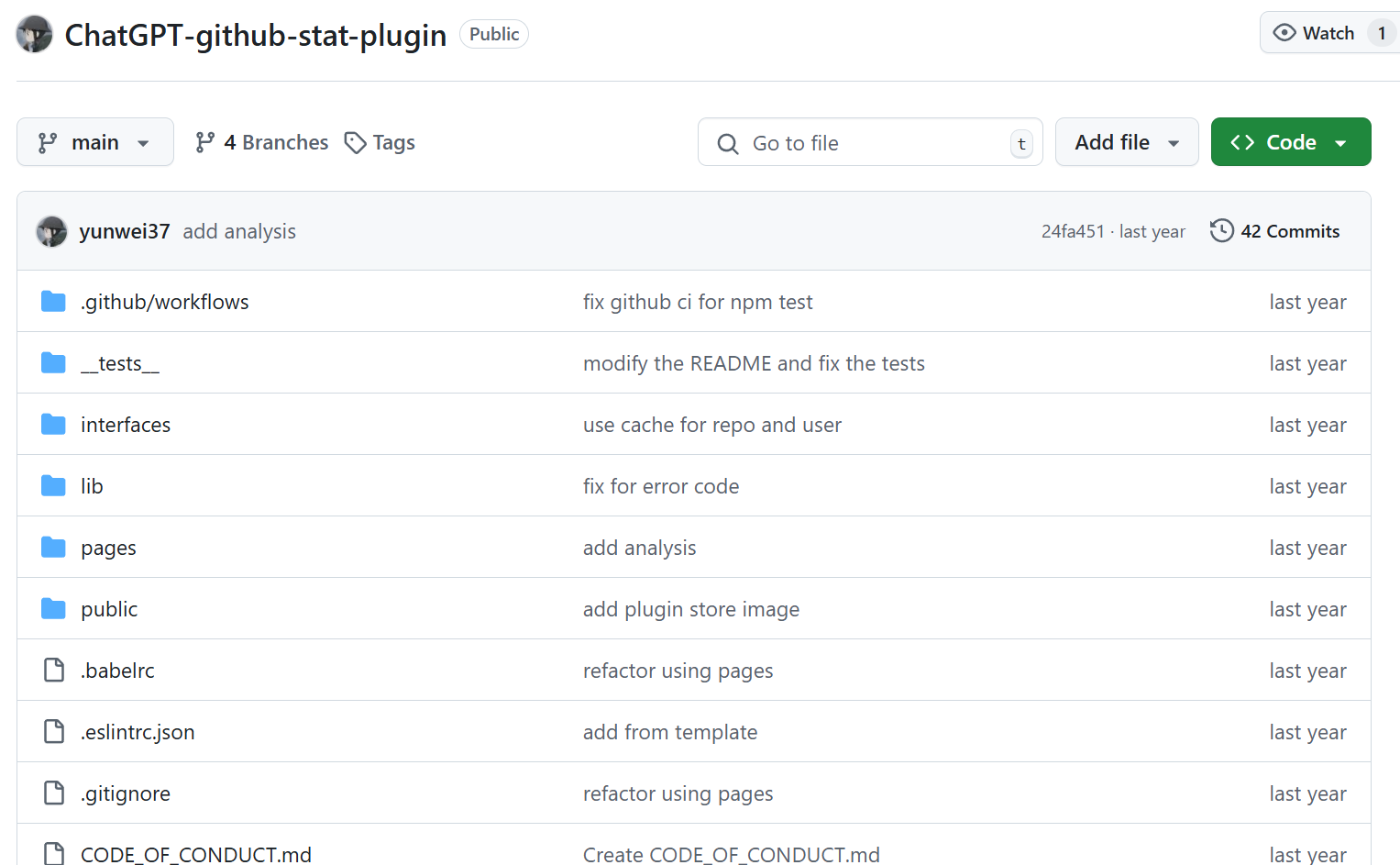}
            \caption{github website}
        \end{subfigure} \\
        \begin{subfigure}{0.4\textwidth}
            \centering
            \includegraphics[width=\linewidth]{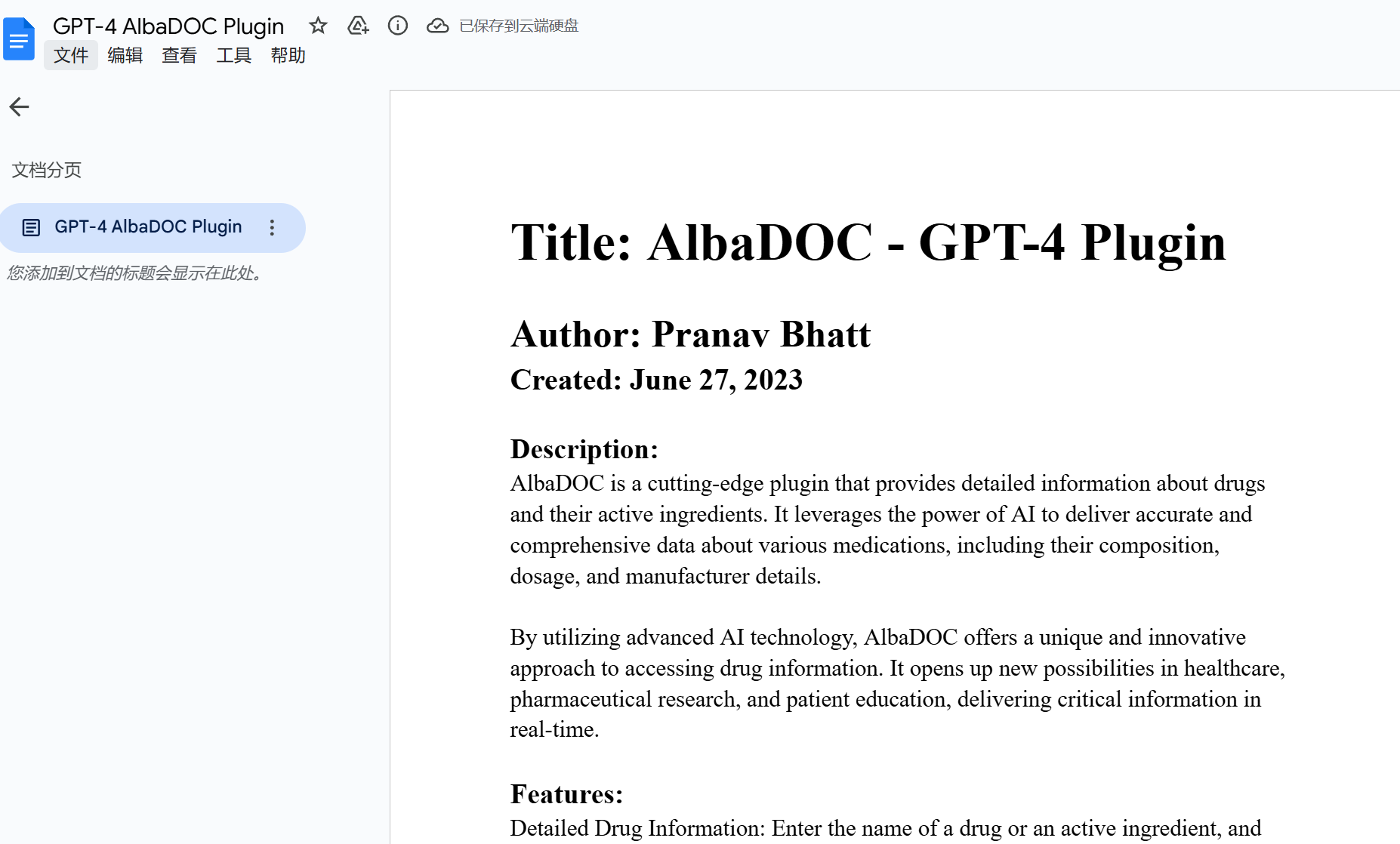}
            \caption{google doc website}
        \end{subfigure} &
        \begin{subfigure}{0.4\textwidth}
            \centering
            \includegraphics[width=\linewidth]{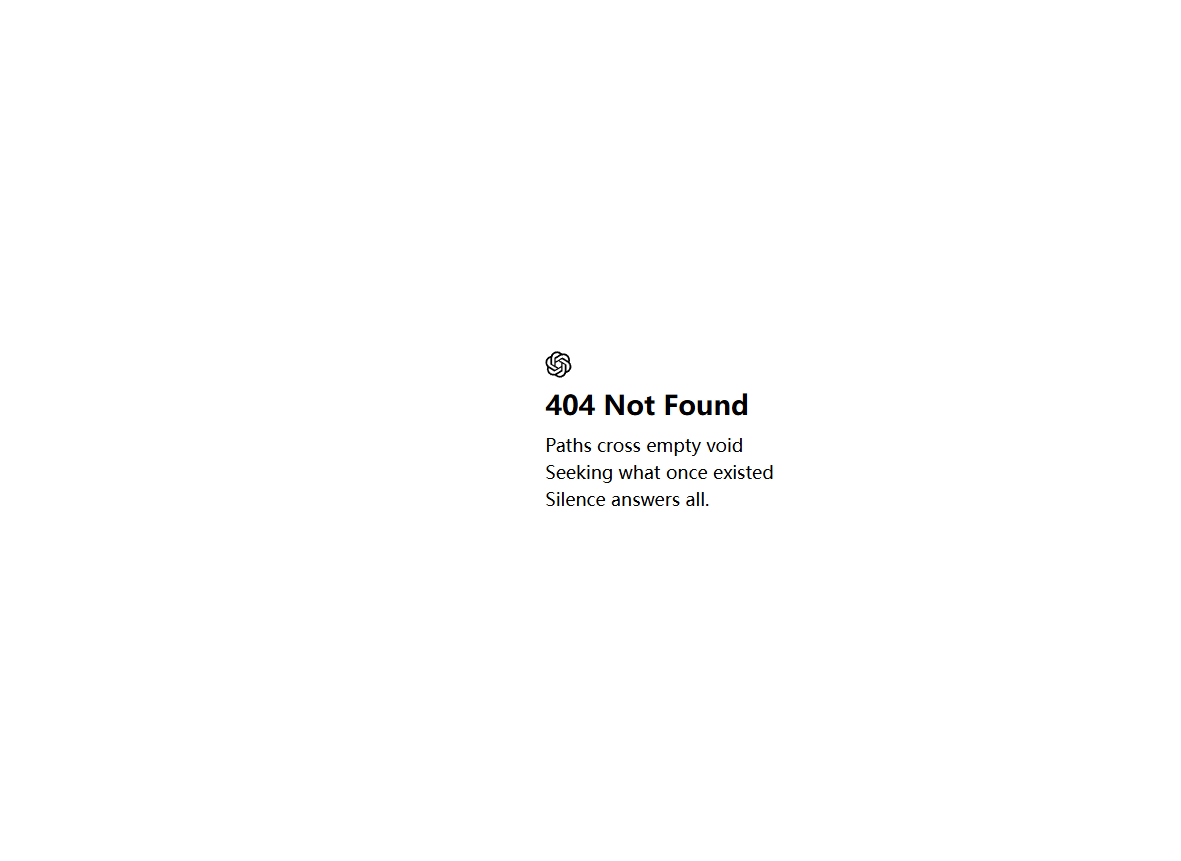}
            \caption{openai website}
        \end{subfigure}
    \end{tabular}
    \caption{Limitations Due to Missing or Invalid External Website Links}
\end{figure}

% 4. 插件开发者未提供有效外部网站的现象
% 在检测过程中，我们发现部分插件的外部网站链接不可访问，这在一定程度上可以视为一种保护措施。具体而言，某些插件的链接因无效或访问限制而无法直接获取 Manifest 文件，反映出这些插件的保护水平较高。这些链接情况包括：

% 原生网址不可到达：在某些情况下，插件的原生网址（即未经过处理或重定向的 URL）并不能直接访问。插件的原生 URL 指向内部服务器、受限资源或并非直接公开的端点，因此这些 URL 无法被外部用户直接到达。
% 开发者通过对资源呈现访问控制，以确保仅授权的应用或用户才能访问资源。
% 在实践中，自动化爬取工具或未经身份验证的请求无法直接访问这些 URL。

%有效网址重定向： 某些插件的开发者将manifest网址进行了重定向
% GitHub 地址：某些插件的开发者仅提供了 GitHub 作为插件的外部链接，而 GitHub 的文件管理规则决定了这些地址通常不会提供明确的 manifest 文件路径。这使得我们无法通过这些 GitHub 地址定位到有效的 manifest 文件，进而影响了我们对这些插件进行信息泄露风险评估的能力。

% OpenAI 保护的域名：我们还发现，部分插件使用了位于 OpenAI 保护下的域名。例如 https://chat.openai.com/dominick.codes 这一类地址，由于受到 OpenAI 的保护，即使我们尝试访问也会返回 403 或 404 错误代码。这意味着这些插件的 manifest 文件在客观上是不可访问的，虽然这不是开发者有意为之，但这类情况无疑限制了外部人员对其进行安全测试的能力。

% Google Doc 链接：例如 https://drive.google.com/file/d/1y39BSFBQsmFFjbqTNVBGB7dGuK0d2lsh/view?usp=sharing 这样的 Google Doc 链接，属于典型的谷歌文档地址。这类地址通常难以通过常规解析方法获取 manifest 文件，并且由于其托管在第三方平台上，外部人员很难对其进行进一步的安全分析。

%这些现象表明，部分插件采取了额外的安全措施，使未经授权的用户或爬虫难以直接访问其 Manifest 文件，从而在一定程度上降低了信息泄露的风险。

During our testing, we found that some plugins had inaccessible links to external websites, which can be considered a protection measure to some extent. Specifically, the links to certain plugins were not directly accessible to the Manifest file due to invalid or restricted access, reflecting a higher level of protection for these plugins. These link situations include:\\

\textbf{Native URLs are unreachable:} In some cases, the plugin's native URLs (i.e., URLs that have not been processed or redirected) are not directly accessible. The plugin's native URLs point to internal servers, restricted resources, or endpoints that are not directly public, so these URLs are not directly reachable by external users.

Developers present access controls to resources to ensure that only authorised applications or users can access them.

In practice, these URLs are not directly accessible by automated crawlers or unauthenticated requests.\\

\textbf{Hidden redirects:} Some plugin developers redirect the URL of manifest files in order to hide them from unauthorised access. This redirection mechanism may result in the file not being directly located or accessed. In this case, the manifest file is designed to be visible only to authorised and authenticated users. Typically, the server checks the visitor's identity and authorisation information, and if the request does not pass authentication, it redirects to an irrelevant or invalid page, preventing unauthorised access to the file.

This authorisation check may include validation of the visitor's authentication information, the originating IP address, the user's proxy, and so on, so that even if an automated crawler is able to trigger the initial URL request, the unauthenticated request will be redirected so that the actual file cannot be located.

The use of this mechanism makes access to manifest files more secure, but also makes it harder to retrieve them.\\

\textbf{GitHub addresses:} Some plugin developers only provided GitHub as an external link to the plugin, and GitHub's file management rules dictate that these addresses usually do not provide an explicit manifest file path. This makes it impossible to locate a valid manifest file via these GitHub addresses, which in turn affects our ability to assess the risk of information leakage for these plugins.\\

\textbf{OpenAI-protected domains:} We also found that some plugins use domains that are under OpenAI protection. For example, addresses such as "https://chat.openai.\\com/dominick.codes" are protected by OpenAI and returned 403 or 404 error codes when we tried to access them. This means that the manifest files of these plugins are objectively inaccessible, which is not intentional on the part of the developers, but it certainly limits the ability of outsiders to test their security.\\

\textbf{Google Doc links:} Google Doc links such as https://drive.google.com/file are typical Google Docs addresses. These addresses are often difficult to obtain manifest files through conventional parsing methods, and because they are hosted on third-party platforms, they are difficult for outsiders to further analyse for security.\\

These phenomena indicate that some plugins have taken additional security measures to make it difficult for unauthorised users or crawlers to directly access their Manifest file, thus reducing the risk of information leakage to a certain extent.

% \subsubsection{Other irregular sites}--要讲在最前面讲了
% 5. 其他不规范的网址
% 除了上述情况，还有一些插件的外部网址结构较为复杂，导致我们无法有效解析出 manifest 文件。常见的情况包括：

% 法律类子网址：一些插件开发者将 manifest 文件托管在法律相关的子目录下，如 legal、privacy、terms 等。这些子目录通常不是常规的 manifest 文件存放路径，因此我们无法通过这些地址找到有效的 manifest 文件。

% 乱码域名：例如 https://xn--pckua2a7gp15o89zb.com/%E5%88%A9%E7%94%A8%E8%A6%8F%E7%B4%84 这样的乱码域名，由于其网址结构的复杂性和不规则性，难以解析出有意义的主网址，进而无法找到 manifest 文件。

% 其他深层子网址：一些插件使用了过于深层的子目录，这些目录不符合一般的网址规范，无法通过常规的解析方式进行访问。

% 针对以上这些情况，我们尝试通过移除特定的关键词后缀（如 legal、terms、tos、p rivacy 等）来探查这些插件的 manifest 文件位置，尽可能穷尽所有可能的路径组合。

%评估没写
%15页
\section{Results and Discussion in API Request Analysis Layer}
%第二层，api request 层
%token有效性的分类和解释  ok
%结果分析
%case study --泄露了什么
%具体危害

%———————————————————————————————————————————————————————————————————————————————%

\subsection{Token Types and Their Impact}
\paragraph{Token}%有效性讨论，我们要解释，基于这个token有效性的问题，需要token，但是token无效，
% 有效性定义
% 在我们的分析中，我们将定义**有效性**的概念，用于描述 API 请求的有效性，它取决于是否需要提供 token 以及提供的 token 是否有效。具体地，我们引入了函数 \(O(T_r, T_v)\)  

In our analysis, we will define the notion of validity for describing the validity of an API request, which depends on whether a token needs to be provided and whether the provided token is valid. Specifically, we introduce the function \(O(T_r, T_v)\)  
token required, token valid, token not required, token not valid, token not required but token valid
\[
T_r = 
\begin{cases} 
1, & \text{API requires a token for access} \\
0, & \text{API does not require a token for access}
\end{cases}
\]

\[
T_v = 
\begin{cases} 
1, & \text{the token is valid} \\
0, & \text{the token is invalid}
\end{cases}
\]

The function \(O(T_r, T_v)\) is defined as follows:
\[
O(T_r, T_v) = 
\begin{cases} 
1, & \parbox[t]{10cm}{API request is successful and \\ returns valid data} \\
0, & \parbox[t]{10cm}{API request fails}
\end{cases}
\]

% 情况 1：需要 token 且 token 有效，能到达
% 描述: 外部请求携带了有效的令牌，API 返回数据。
% 问题: 在此情况下，系统误将未经授权的外部请求视为合法请求，导致敏感数据泄露。
% 安全隐患: 这种情况表明存在漏洞，未充分隔离内部授权机制，使外部非法请求在提供有效令牌后也能获得访问权限。
\paragraph{Case 1}: token required and token is valid
\[
T_r = 1, \quad T_v = 1, \quad O(T_r, T_v) = 1
\]
Meaning: The API requires authentication and provides a valid token, the API request succeeds and returns valid data.\\
Practical impact: In this case, the system mistakenly treats unauthorised external requests as legitimate ones, leading to the leakage of sensitive data.\\
% Security significance: This situation indicates a vulnerability that does not adequately isolate the internal authorisation mechanism, allowing an illegal external request to gain access even after a valid token has been provided.

% 情况 2：需要 token 但 token 无效,可以直接访问API，不能到达

% 描述: 外部请求无论是否提供有效令牌，均无法获得数据。
% 安全性: 这种机制确保了外部请求不会访问到敏感数据，起到了良好的保护作用。
% 安全意义: 表明API设计符合安全性预期，确保未经授权的外部请求无法访问数据，减少了泄露风险。
\paragraph{Case 2}: token is required but invalid, you can access the API directly.
\[
T_r = 1, \quad T_v = 0/1, \quad O(T_r, T_v) = 0
\]
Meaning: API requires authentication, but even we add token or not, the API request fails and no valid data is returned.\\
Impact: This mechanism ensures that external requests do not access sensitive data, providing good protection.\\
% Security implications: Indicates that the API design meets security expectations, ensuring that unauthorised external requests do not access the data, reducing the risk of compromise.

%case3 
\paragraph{Case 3}: Token is required, and the token is invalid, you can access the API directly.

% 描述: 外部请求提供了无效令牌，但API仍然返回数据。
% 问题: 在此情况下，API 似乎忽略了令牌的有效性检查，导致数据泄露。
% 安全隐患: 这种情况揭示了API授权机制的严重缺陷，意味着即使没有有效令牌，外部请求也能获得数据，从而给系统带来高风险。

\[
T_r = 1, \quad T_v = 0, \quad O(T_r, T_v) = 1
\]
Meaning: The API does require authentication, but a invalid token is provided (though not necessary), the API request succeeds and returns valid data without token.
Impact: In this case, the API ignore token validity checks, leading to data leakage.\\

% Security implications: This situation reveals a serious flaw in the API authorisation mechanism, meaning that external requests can obtain data even without a valid token, thus posing a high risk to the system.

% 情况 3：不需要 token,可以直接访问API
% 含义：API 不需要身份验证，因此无论 token 是否有效，API 请求都会成功。
% 实际影响：这是开放式 API 的典型场景，任何用户都可以访问，无需身份验证。
% 安全意义：虽然方便了用户，但开放 API 可能带来安全隐患，尤其是当 API 处理敏感数据时。
\paragraph{Case 4}: No token is required and whether the token is valid or invalid, you can access the API directly
\[
T_r = 0, \quad T_v = 1/0, \quad O(T_r, T_v) = 1
\]
Meaning: The API does not require authentication, so the API request will succeed regardless of whether the token is valid or not.\\
Practical implications: this is a typical scenario for an open API, which can be accessed by any user without authentication.\\
% Security Implications: Although it is convenient for users, an open API may bring security risks, especially when the API handles sensitive data.

% 情况 4：不需要 token 且 token 无效，无法访问API
% 含义: API 不需要验证令牌，因此即使请求中提供了无效令牌，API 也不会返回数据。
% 实际影响: 此情况表明API具备良好的鉴别能力，能够正确拒绝未经授权的请求，不向外部请求暴露任何数据。
% 安全意义: 这种设计显示出良好的安全实践，对于无需授权的API，严格控制数据访问，避免外部非法请求获取敏感信息。
\paragraph{Case 5}: token not required and whether the token is valid or invalid, API cannot be accessed
\[
T_r = 0, \quad T_v = 0/1, \quad O(T_r, T_v) = 0
\]
Meaning: The API does not require a token, so even if an token is provided or not, the API is still not able to return valid data.\\
Practical implications: This scenario shows that the API has good forensic capabilities to correctly reject unauthorised requests without exposing any data to external requests.\\
% Security implications: This design shows good security practices for APIs that do not require authorisation, strictly controlling access to data and avoiding illegal external requests for sensitive information.

\begin{table}[h]
\centering
\caption{Analysis of API Request Cases and Security Implications}
\renewcommand{\arraystretch}{1.8} % Increase line height specifically for the header
\begin{tabular}{|p{1.3cm}|p{1.2cm}|p{2.5cm}|p{2.5cm}|p{2.5cm}|p{4cm}|}
\hhline{|======|} % Double line at top
\textbf{Case} & \textbf{Count} & \textbf{Token Requirement (Tr)} & \textbf{Token Validity (Tv)} & \textbf{Request Outcome (O)} & \textbf{Security Implications} \\
\hhline{|======|} % Double line under header
Case 1 & 8 & Required & Valid & Success & High risk of data leakage \\
\hline
Case 2 & 74 & Required & Valid/Invalid & Failure & Effective access protection \\
\hline
Case 3 & 24 & Required & Invalid & Success & Significant authorization flaw \\
\hline
Case 4 & 141 & Not Required & Valid/Invalid & Success & Risk in open access \\
\hline
Case 5 & 98 & Not Required & Invalid & Failure & Strong data protection \\
\hhline{|======|} % Double line at bottom
\end{tabular}
\end{table}

\subsection{API Request Success and Failure Analysis}
%我们通过python脚本伪装成合法请求，来探查api问题。为了保证结果的鲁棒性，我们通过更换IP和服务器进行了多次探查，最后得到相关结果，在本节，我们将基于得到的结果，分析5个典型案例的构成和成因，并讨论其可能导致的安全问题和潜在危害。
%基于manifest泄露的368个插件上，我们剔除了manifest文件本身的不规范的部分。最终，，有173个插件在我们伪装成外部请求的情况下成功返回了有效的响应。在这173个插件中：

We probe API issues through python scripts disguised as legitimate requests. In order to ensure the robustness of the results, we probed several times by changing IP and servers, and finally obtained the relevant results. In this section, based on the obtained results, we will analyse the composition and causes of 5 typical cases, and discuss the security issues and potential hazards they may lead to.

On the 373 plugins leaked based on the manifest, we eliminated the irregularities in the manifest file itself. In the end,, 173 plugins managed to return valid responses when we disguised them as external requests and 172 failed.After we classify the cases according to our case above, we get the following result:\newline

% Case 1:该情况发生在8个插件中，此类插件要求提供有效的令牌才能访问，但由于未能有效隔离内部授权机制，表明存在漏洞，外部非法请求在提供有效令牌后也能获得访问权限。这表明系统存在漏洞，未能完全防止未经授权的外部访问。

% Case 2: 该情况发生在74个插件中，表明API设计符合安全性预期，确保未经授权的外部请求无法访问数据，减少了泄露风险。然而风险在于，认证机制的缺陷使得即便经过身份验证，安全风险仍然存在。特别是在第三方认证系统自身存在漏洞或与插件集成不完善的情况下，攻击者可能会绕过认证机制获取数据。在这种情况下，主要的泄露风险控制责任从 OpenAI 平台转移到第三方网站，进而增加了泄露的可能性。此外，第三方认证的泄露风险不容忽视，尤其是当认证信息可能被用于恶意用途时。

% Case 3: 该情况显示有24个插件虽然要求令牌，但即使提供无效令牌也能获得访问权限。这种情况揭示了API授权机制的严重缺陷，给系统带来风险。这实际上使令牌失效，即使开发人员设计了令牌机制，仍可能存在漏洞或令牌的使用不当。开发人员可能因此误以为系统安全，而实际情况却并非如此。

% Case 4:这是最多的一类，有141个插件， 表明这些插件为开放式API，不要求令牌验证，也从而允许外部请求直接访问数据。虽然方便了用户，但开放 API 可能带来安全隐患，尤其是当 API 处理敏感数据时。开发者可能低估了所传递信息的敏感性，认为插件传递的信息无需加密或身份验证。然而，这种判断有时会导致严重的信息泄露风险。例如，看似无关紧要的信息可能会在与其他数据结合后，暴露出用户隐私或系统配置等敏感内容。

% Case 5 : 该情况显示有24个插件的API没有令牌需求，但也不会响应外部请求。这种设计在98个插件中实现了严格的访问控制，显示出良好的安全实践，防止了未经授权的访问，有效避免了敏感数据的泄露风险。

\textbf{Case 1:} This occurred in 8 plugins that required a valid token for access, but the failure to effectively segregate the internal authorisation mechanism suggests a vulnerability whereby unauthorised external requests were able to gain access even after a valid token had been provided. This indicates a vulnerability in the system that does not fully prevent unauthorised external access.

\textbf{Case 2:} This occurred in 74 plugins, indicating that the API design meets security expectations and ensures that unauthorised external requests cannot access the data, reducing the risk of compromise. The risk, however, is that flaws in the authentication mechanism allow security risks to remain even after authentication. Especially if the third-party authentication system itself has vulnerabilities or is poorly integrated with the plugin, an attacker may be able to bypass the authentication mechanism to gain access to the data. 

In this case, the main responsibility for leakage risk control is shifted from the OpenAI platform to the third-party website, which increases the likelihood of leakage. Additionally, the risk of compromise of third-party authentication cannot be ignored, especially if the authentication information could be used for malicious purposes.

\textbf{Case 3:} This scenario shows 24 plugins that require a token but are able to gain access even if an invalid token is provided. This case reveals a serious flaw in the API authorisation mechanism that poses a risk to the system. It effectively invalidates the token, and even if the developer has designed the token mechanism, there may still be vulnerabilities or improper use of the token. The developer may therefore mistakenly believe that the system is secure, when in fact it is not.

\textbf{Case 4:} This is the largest category, with 141 plugins, indicating that these plugins are open APIs that do not require token authentication and thus allow external requests to access data directly. While convenient for users, open APIs can pose a security risk, especially if the API handles sensitive data. Developers may underestimate the sensitivity of the information being passed, believing that the information being passed by the plugin does not need to be encrypted or authenticated. However, this judgement can sometimes lead to serious risks of information leakage. For example, seemingly inconsequential information may, when combined with other data, reveal sensitive content such as user privacy or system configuration.

\textbf{Case 5:} This design implements strict access control in 98 plugins, showing good security practices, preventing unauthorised access and effectively avoiding the risk of sensitive data leakage.

\subsection{Analysis of Failed Requests and Their Causes}

%未能检测到有效请求的插件共计159个，占比47.9%，这类请求显示出了不同的危害性原因，我们按照他错误的原因进行了主要的分类，其中包括:
% 缺少授权（Lack Authorization）：55个请求失败，因为请求缺少必要的授权信息，无法通过身份验证。
% 客户端错误（Client Errors）：62个请求由于客户端的错误（如请求格式错误或其他客户端问题）导致失败。
% 限流（Rate Limiting）：42个请求因为触发了API的限流机制而失败，即由于请求频率过高或者检测到了脚本，API拒绝提供服务。

%如果我们按照token的情况分类，那我们可以将不可达到情况，分成有token和无token，这两种case体现了插件开发商良好的保护措施。由此可见，虽然部分插件对token或验证机制有较严格的要求，体现了开发商在保护用户数据方面的良好实践，但仍有一部分插件低估了潜在的安全风险，忽视了信息的敏感性，给攻击者提供了可乘之机。

The total number of plugins that failed to detect a valid request was 173(50.19\%). These requests showed different reasons for their harmful effects, and we have classified them according to the main reasons for their errors, which include.

Lack Authorization: 58 requests failed because they lacked the necessary authorisation information to pass authentication.

Client Errors: 66 requests failed due to client-side errors such as incorrectly formatted requests or other client-side issues.

Rate Limiting: 49 requests failed because the API's rate limiting mechanism was triggered, i.e., the API refused to serve the request because it was requested too often or a script was detected.\\

\begin{table}[h!]
    \centering
    \caption{Failed Requests and Different Token Types Influence }
    \begin{tabular}{@{}llc@{}}
        \toprule
        \textbf{Category} & \textbf{Count} & \textbf{Percentage} \\ \midrule
        Can Retrieve Valid Information & 173 & 52.1\%\\ \\ \hline
        \quad With No Token & 141 &  \\
        \quad With Authentication Requirements & 27 &  \\
        \quad With Verification Token & 5 &  \\ \hline
        Fail to Retrieve Valid Information & 159 & 47.9\%\\ \\ \hline
        \quad Lack Authorization & 58 &  \\
        \quad Client Errors & 66 &  \\
        \quad Rate Limiting & 49 &  \\ 
        \bottomrule
    \end{tabular}
\end{table}

\subsection{Analysis of Different Token Types in Response to illegal Requests}
%我们将进一步探讨插件的身份验证机制及其对安全性的影响。通过对不同类型Token的分析，无需Token访问、需要身份验证的插件以及需要验证Token是chatgpt插件目前存在的三种token形式，本节评估了它们的泄露频次，结果如下：

%无需Token即可成功请求：在我们的研究中，共有141个插件允许用户在不需要身份验证的情况下成功获取信息。
                                      
%有身份验证要求：有5个插件要求身份验证才能成功访问。这里的身份验证要求是指来自第三方机构的验证重定向，比如基于谷歌的账号验证。这些插件要求用户在其他网站上拥有账号，才能进行相关操作。但是在此情况下，第三方验证机制未能有效发挥其应有的作用。

%需要验证Token（With Verification Token），27个需要token验证的插件成功，这类包括使用http/json方式，配合唯一标识码验证。此类受manifest文件泄露影响较深，当manifest文件泄露了，这个也会泄露。

% % 并且，通过我们的监控发现，在manifest文档里的唯一标识码是一个长期唯一标识码，即他并不会随着时间和对话的改变而更新。这进一步增加泄露和监听风险。该类危害在于，这种唯一标识码如果被攻击者获取到，很容易用于重放攻击，允许恶意用户多次模拟合法请求，进而访问到敏感信息。

% % 这增加了恶意监听的可能性，尤其是在信息流被劫持或泄露的情况下，攻击者可以通过拦截数据包持续获取未加密或未保护的用户数据。

% 我们进一步统计了各Token在ChatGPT插件商店中的应用广泛性和有效性，结果如下：

% 无须 Token 类别： 共有 239 个插件不需要身份验证，约占总数的 82.1%。其中，成功访问的插件为 141 个，失败的插件为 98 个，成功率为 58.9%。尽管这些插件无需身份验证，但开放访问的设计可能导致敏感数据的泄露。

% OAuth 类别： 共有 70 个插件使用 OAuth 身份验证，占总数的 24.4%。成功访问的插件为 27 个，失败的插件为 43 个，成功率为 38.6%。这表明，尽管OAuth作为一种身份验证机制被广泛使用，但其成功率较低，显示出潜在的安全隐患。

% Bearer Token 类别： 共有 34 个插件使用 Bearer Token，占总数的 11.8%。成功访问的插件为 5 个，失败的插件为 29 个，成功率为 14.7%。该类别的插件成功率极低，显示出Bearer Token在安全性方面的不足，可能导致数据泄露风险的增加。

% 这些统计结果表明，无需Token类别依然是最为广泛的授权类型，然而这种类型的泄露几率最高；OAuth使用相对较少，但其泄露成功率高于Bearer Token，后者的泄露成功率最低。通过对这些数据的分析，我们可以更好地理解不同身份验证机制在安全性方面的表现及其潜在的改进方向。

In this part, We will further explore the authentication mechanisms of plugins and their impact on security. By analysing different types of Token, access without Token, plugins requiring authentication, and requiring verification of Token are the three forms of token that currently exist in chatgpt plugins, and this section evaluates their leakage frequency with the following results:

\paragraph{Request without Token:} In our study, a total of 141 plugins allowed users to successfully access information without requiring authentication.
                                      
\paragraph{Request With Verification Token:} 27 plugins that require token verification were successful, this category includes the use of http/json method with unique identifier code verification. This category is deeply affected by manifest file leaks, when the manifest file leaks, this will also leak.

And, through our monitoring, we found that the unique identifier in the manifest document is a long-term unique identifier, i.e., it will not be updated as time and dialogue change. This further increases the risk of leakage and eavesdropping. The danger is that this unique identifier, if obtained by an attacker, can easily be used in a replay attack, allowing a malicious user to simulate a legitimate request multiple times and thus access sensitive information.

This increases the likelihood of malicious eavesdropping, especially if the information flow is hijacked or compromised, and attackers can persistently access unencrypted or unprotected user data by intercepting packets.

\paragraph{Request With Authentication Requirement:} 5 plugins required authentication for successful access. Authentication requirements here are authentication redirects from third-party organisations, such as Google-based account verification. These plugins require the user to have an account on another website in order to perform the relevant action. But in this case, the third-party authentication mechanism fails to work effectively as it should.\\

We further counted the breadth and effectiveness of each Token in the ChatGPT plugin shop, and the results are as follows:

\paragraph{Request without Token:}  A total of 239 plugins do not require authentication, accounting for about 82.1\% of the total. Of these, 141 plugins were successfully accessed and 98 failed, a success rate of 58.9\%. Although these plugins do not require authentication, the open-access design may result in the disclosure of sensitive data.

\paragraph{Request With Verification Token:} A total of 70 plugins, or 24.4\% of the total, used OAuth authentication. Twenty-seven plugins were successfully accessed and 43 failed, for a success rate of 38.6\%. This indicates that although OAuth is widely used as an authentication mechanism, its success rate is low, showing potential security risks.

\paragraph{Request With Authentication Requirement:}  A total of 34 plugins used Bearer Token, accounting for 11.8\% of the total. The number of plugins that were successfully accessed was 5, and the number of plugins that failed was 29, a success rate of 14.7\%. The extremely low success rate of plugins in this category shows the lack of security of Bearer Token, which may lead to an increased risk of data leakage.\\

These statistics show that the No Token Required category is still the most widely used authorisation type, however, this type has the highest chance of leakage; OAuth is used relatively less, but its leakage success rate is higher than Bearer Token, which has the lowest leakage success rate. By analysing this data, we can better understand the performance of different authentication mechanisms in terms of security and their potential directions for improvement.
    \begin{figure}
    \centering
    \includegraphics[width=0.75\linewidth]{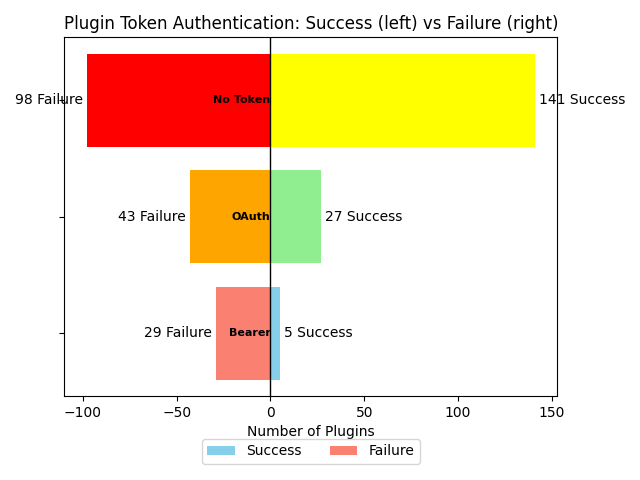}
    \caption{Plugin Token Authentication}
    \label{fig:enter-label}
\end{figure}

\subsection{Case Study 1: Severity Analysis of Unauthorized Requests in OAuth-Authorized Plugins}
% 为了更有效地评估 API 外部泄露的安全影响，本节将进行案例研究。在 layer 2 的基础上，我们将重点分析在 OAuth token 下，ChatGPT 插件所要求的用户授权范围，并评估可能泄露的严重性。通过这个案例，我们期望深入了解 ChatGPT 插件涉及的具体权限范围，从而更全面地评估其潜在风险。

In order to more effectively assess the security impact of external API leakage, this section will conduct a case study. Based on Layer 2, we will focus on analysing the scope of user authorisations required by the ChatGPT plugin under OAuth token and assess the severity of possible leakage. Through this case study, we expect to gain a deeper understanding of the specific scope of permissions involved in the ChatGPT plugin, and thus assess its potential risks more comprehensively.
% 
    % \\% 本案例研究的目标是通过对OAuth权限（scope字段）的分析，识别潜在的权限泄露风险。我们通过GPT-4模型的自然语言处理能力，对权限描述进行自动化总结，并评估每类权限的安全性。这将帮助我们更好地理解每个scope字段所代表的权限，并发现可能导致权限滥用或数据泄露的风险。具体目标如下：
    % 数据收集：从Layer 2的实际OAuth授权API中提取scope字段的权限描述，涵盖插件或应用可访问的资源类型和操作权限。
    
    % 应用GPT-4模型：利用TF-IDF模型解析权限描述，将其化为向量，由于缺乏其他先验数据，我们使用无监督学习，通过计算名称的余弦相似度将相似权限进行归类。
    
    % 权限分类与总结：通过模型自动将权限分类为全局访问、读写权限、身份信息权限等，并提供简明的功能解释。
    
    % 风险分析：基于生成的总结，深入分析各类权限的潜在风险，例如，全局访问权限可能导致更高的数据泄露风险
    % 提出改进建议：根据分析结果，提出改进权限管理的建议，帮助减少权限泄露的风险。
    
\subsubsection{Objectives and Methodology:}
The goal of this case study is to identify potential risks of privilege leakage by analysing OAuth permissions (scope fields). We automate the summarisation of permission descriptions and evaluate the security of each type of permission through the natural language processing capabilities of the TF-IDF model. This will help us to better understand the permissions represented by each scope field and identify risks that may lead to permissions misuse or data leakage. The specific objectives are as follows:

\begin{itemize}
    \item \textbf{Data collection:} Extract permissions descriptions for scope fields from the actual OAuth authorisation API of Layer 2, covering the types of resources and operation permissions accessible by the plugin or application.

    \item \textbf{Application of TF-IDF model:} Parse permission descriptions into vectors using the TF-IDF model. Due to the lack of other a priori data, we use unsupervised learning to categorize similar permissions by calculating the cosine similarity of the names.

    \item \textbf{Permission classification and summary:} The model automatically classifies permissions into global access, read/write permissions, identity information permissions, etc., and provides concise functional explanations.

    \item \textbf{Risk analysis:} Based on the generated summary, analyze the potential risks of each type of permission in depth, for example, global access permissions may lead to a higher risk of data leakage.

\end{itemize}

% \begin{figure}
%     \centering
%     \includegraphics[width=1\linewidth]{fig/code to find feature of scope.png}
%     \caption{Code to find feature of scope}
%     \label{fig:enter-label}
% \end{figure}

    \subsubsection {Visualization of Permission Leakage}

 In addition to the list of plugins that do not explicitly mark the scope of scope permissions or have an empty scope (47.9\%), we counted the assignments of explicitly marked permissions and classified them into six categories based on the nature of the permissions:

\paragraph{Global/broad access permissions}
These permissions allow the plugin to perform full operations on the API, or access a large number of resources, and carry a high risk of permissions.

If the token is compromised, an attacker may be able to access all API functionality, leading to widespread data leakage and tampering.

For example, the all-class permission allows the API unrestricted access to all functionality. Another example is baas-full-access (1.408\%), which allows full access to backend services.

Due to the broad coverage of these permissions, an attacker can perform a large number of unauthorised operations, which could lead to a serious system security breach.

Global access permissions account for 12.675\% of all OAuth class permissions.

\paragraph{Read-write permission}
Read-write privileges allow reading and modifying operations on specific resources, and the risk is relatively concentrated.If compromised, an attacker can read sensitive user data and even modify or delete user information.An attacker could use these permissions to manipulate user data or access information when the user is not online, increasing the potential threat.

In the manifest file of the ChatGPT plugin, these permissions appear as read+write and read write read offline\_access (1.408\%), the latter of which allows not only read and write operations but also offline access to data.

Read and write permissions account for 9.858\% of all OAuth permissions.

\paragraph{OpenAI platform-related permissions}
Some scope fields are related to the OpenAI platform, specifically allowing plugins to access services provided by OpenAI (e.g., ChatGPT).If these permissions are compromised, an attacker could perform unauthorised operations through OpenAI's APIs, potentially negatively impacting the plugin services themselves.

For example, the openai access permission (5.634\%) and the oai11/global permission (1.408\%) allow access to OpenAI's global resources. Another permission, openid email https://openai.videoinsights.io/all (1.408\%), allows access to the OpenAI video insights system.

If these permissions are abused, an attacker could use OpenAI's services for malicious purposes.Permissions related to the OpenAI platform account for 8.459\% of all OAuth-type permissions.

\paragraph{Identity and Email Privileges}
These permissions relate to sensitive data such as user identities, emails, and personal data.If compromised, they can lead to privacy breaches or social engineering attacks.

The permissions included are: email permission (5.634\%), which allows access to the user's email; and profile permission (2.817\%), which allows access to the user's basic information. Other permissions such as w\_member\_social openid profile email (1.408\%), which allows access to social accounts and identity information, and openid offline\_access (1.408\%), which allows access to data when the user is offline. There is also the playlist-modify-public user-read-email permission, which allows modification of playlists and access to users' emails.

Once compromised, these permissions could be used by an attacker to commit identity theft, forge identities, or perform social engineering attacks, impacting personal privacy and platform security.

Identity information and email permissions account for 15.491\% of all OAuth-type permissions.

 \begin{figure}
    \centering
    \includegraphics[width=1\linewidth]{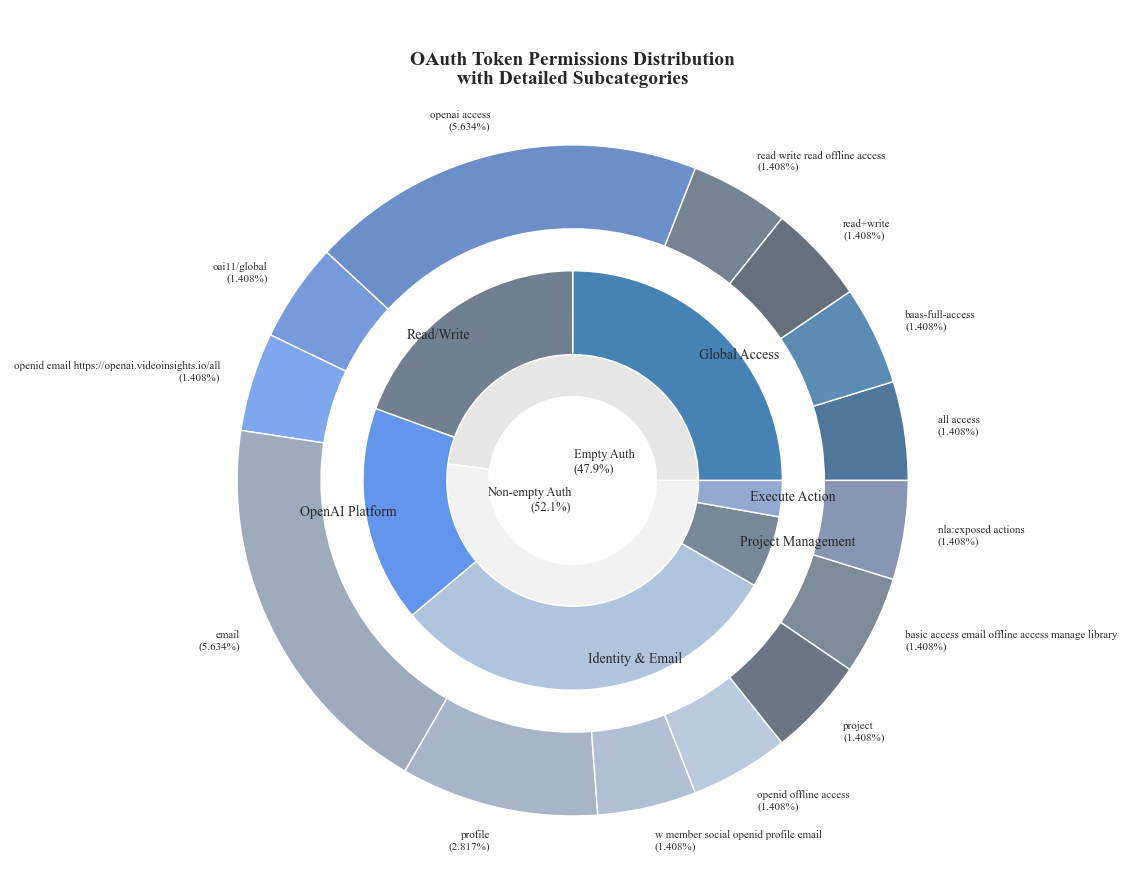}
    \caption{OAuth Token Permissions Distribution Detailed Subcategories}
    \label{fig:enter-label}
\end{figure}

\paragraph{Project and task management permissions}
These scopes allow plugins to manage specific projects or tasks.

If compromised, an attacker may be able to change project data or task details, impacting the organisation's workflow and the security of sensitive projects.A compromise of project and task management privileges could allow an attacker to gain complete control over a project or even tamper with or delete critical data.

This category includes project (1.408\%), which allows access to and modification of project data, and basic\_access email offline\_access manage\_library (1.408\%), which allows the management of user repositories and offline access rights.

Project and task management permissions account for 2.816\% of all OAuth class permissions.

\paragraph{Execute Action Permissions}
These scopes allow plugins to perform specific public operations, usually involving automated tasks.

If compromised, an attacker could use these permissions to perform unauthorised operations that could lead to the execution of malicious automation tasks and compromise the integrity of the system.

For example, the nla:exposed\_actions (1.408\%) permission allows plugins to perform public tasks or commands, which can be abused by an attacker to perform destructive or malicious operations.

The executed\_actions permission accounts for 1.408\% of all OAuth class permissions.

%如果太多，case study就少写点，评估看看往哪里放///还有？？？？和图要插进去
%%一致性---6-10页
%%探查什么
%%case2 就是那个所谓的那啥
%%评估--可能出现的问题
%%
\section{Results and Discussion in Consistency and Integrity Analysis Layer}
\subsection{Definition and Findings of Consistency}
% ？定义与一致性问题的发现
% 我们基于 ChatGPT 的工作流程，从两个方面探讨了插件的一致性问题。首先，ChatGPT 平台在与用户和插件交互时分别使用了 name_for_human 和 name_for_model 两个字段，其中 name_for_human 向用户展示插件的用途和信息，而 name_for_model 则供大语言模型使用。这些字段的不一致可能导致一系列问题。此外，插件还通过 legal_info_url 字段向用户提供法律信息作为法律指引。

% 通过多次脚本测试，我们发现了 368 个可探查的 manifest 文件中有 69 个存在一致性问题，主要表现为模糊不清或频繁更改的插件名称。例如，某些开发者使用相同名称 “MixerBox OnePlayer” 为多达 17 个具有不同用途的插件命名。这种情况可能导致用户在发送请求时，多个插件同时响应，特别是当用户安装了同一开发者提供的多个插件时。此外，有些插件通过在名称中添加或更改首字母（如将 “Digital Pet” 改为 “A Digital Pet”）以增加其在排名中的可见度，获得更高的优先级和推荐位置。

We explored two aspects of plugin consistency based on the workflow of ChatGPT. First, the ChatGPT platform uses two fields, name\_for\_human and name\_for\_model, when interacting with users and plugins respectively, where name\_for\_human shows users the usage and information of the plugin, and name\_for\_model is used by the big language model. 

Inconsistencies in these fields can lead to a number of problems. Additionally, the plugin provides users with legal information as a legal guide via the legal\_info\_url field.

Through multiple script tests, we found consistency issues in 69 out of 373 explorable manifest files, mainly in the form of ambiguous or frequently changing plugin names. 

For example, some developers use the same name ‘MixerBox OnePlayer’ for up to 17 plugins with different purposes. This situation may result in multiple plugins responding at the same time when a user sends a request, especially if the user has installed multiple plugins provided by the same developer.

Additionally, some plugins gain higher priority and recommended positions by adding or changing initials in their names (e.g., changing ‘Digital Pet’ to ‘A Digital Pet’) to increase their visibility in rankings.

\begin{table}[h]
    \centering
    \caption{Inconsistencies in Plugin Metadata}
    \begin{tabular}{@{}lll@{}}
        \toprule
        \textbf{Type of Inconsistency} & \textbf{Number of Plugins} \\ \midrule
        Inconsistent Names              & 34                          \\
        Different Descriptions          & 8                           \\
        Mismatched Legal Document URLs   & 27                          \\ \bottomrule
    \end{tabular}
\end{table}

\subsection{Potential Issues Caused by Inconsistency}
% 一致性问题可能引发的潜在问题
% 插件元数据中的不一致性可能导致多个方面的问题，影响用户体验、安全性以及平台的完整性：

% 误导性操作

% 用户面对的 name_for_human 与模型使用的 name_for_model 字段存在差异，这种不一致可能导致用户对插件功能产生误解，从而在不知情的情况下进行错误的操作。例如，用户可能因误解插件的功能而授予不必要的权限，增加了意外操作和权限滥用的风险。
% 恶意伪装

% 一些开发者可能利用 name_for_human 字段进行欺骗性描述，从而掩盖插件的真实意图，诱导用户认为插件是合法且安全的。这种操作可以绕过用户的安全感知，访问敏感数据，造成潜在的安全漏洞。通过伪装，插件能够在用户不知情的情况下执行恶意行为，进一步影响平台的安全性。
% 权限和模型滥用

% 由于 name_for_human 和 name_for_model 不一致，插件可能在平台和用户之间隐藏其真实意图，导致模型调用错误。例如，一些冷门插件在描述或字段中加入热门用户的名称或关键词，以此增加点击率，导致模型错误调用。此外，由于用户对插件权限的理解可能基于字段内容不充分，用户可能错误地授予过多或非必要的权限，增加了隐私泄露的风险。
% 法律信息的误导性

% legal_info_url 字段中的法律信息用于提供给用户的法律指引，但由于信息不对称性，一些插件提供的法律信息可能存在误导性。这种情况容易导致用户对开发者的责任理解不清，甚至可能出现恶意规避责任的现象，影响用户对插件的知情同意水平。

Inconsistencies in plugin metadata can lead to problems in several areas, affecting user experience, security, and the integrity of the platform:

\paragraph{Misleading operations}

There is a discrepancy between the name\_for\_human that the user is confronted with and the name\_for\_model field used by the model, and this inconsistency can lead to users misunderstanding the plugin's functionality and unknowingly performing incorrect operations. 

For example, a user may grant unnecessary permissions due to a misunderstanding of the plugin's functionality, increasing the risk of unintended operations and permissions abuse.

\paragraph{Malicious spoofing}

Some developers may utilise the name\_for\_human field for deceptive descriptions, thereby masking the true intent of the plugin and inducing users to believe that the plugin is legitimate and secure. 

This manipulation can bypass the user's security perceptions and access sensitive data, resulting in a potential security breach. Through camouflage, the plugin is able to execute malicious behaviours without the user's knowledge, further compromising the security of the platform.

\paragraph{Privilege and model abuse}

Due to inconsistencies between name\_for\_human and name\_for\_model, plugins may hide their true intentions between the platform and the user, resulting in incorrect model calls. For example, some cold plugins add the names or keywords of popular users to descriptions or fields as a way to increase click-through rates, leading to incorrect model calls.

In addition, because the user's understanding of plugin permissions may be inadequate based on field content, the user may incorrectly grant too many or non-essential permissions, increasing the risk of privacy breaches.

\paragraph{Misleading legal information}

legal\_info\_url: The legal information in the legal\_info\_url field is used to provide legal guidance to users, but due to information asymmetry, the legal information provided by some plugins may be misleading. This situation can easily lead to a lack of understanding of the developer's responsibilities, and may even lead to malicious circumvention of responsibilities, affecting the user's level of informed consent to the plugin.

\subsection{Case Study 2: Examples of Inconsistency in Path Exploration}

\begin{figure}[H]
    \centering
    \includegraphics[width=1\linewidth]{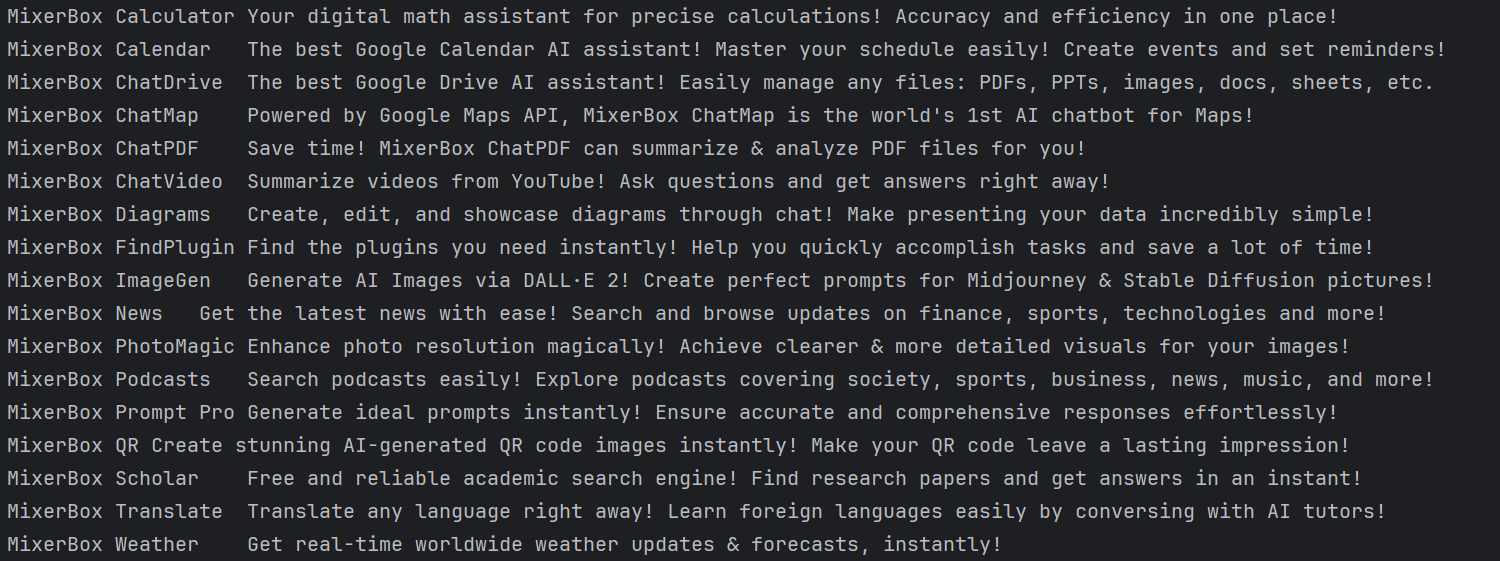}
    \caption{MixerBox plugins list}
    \label{fig:enter-label}
\end{figure}

\begin{figure}[H]
    \centering
    \includegraphics[width=1\linewidth]{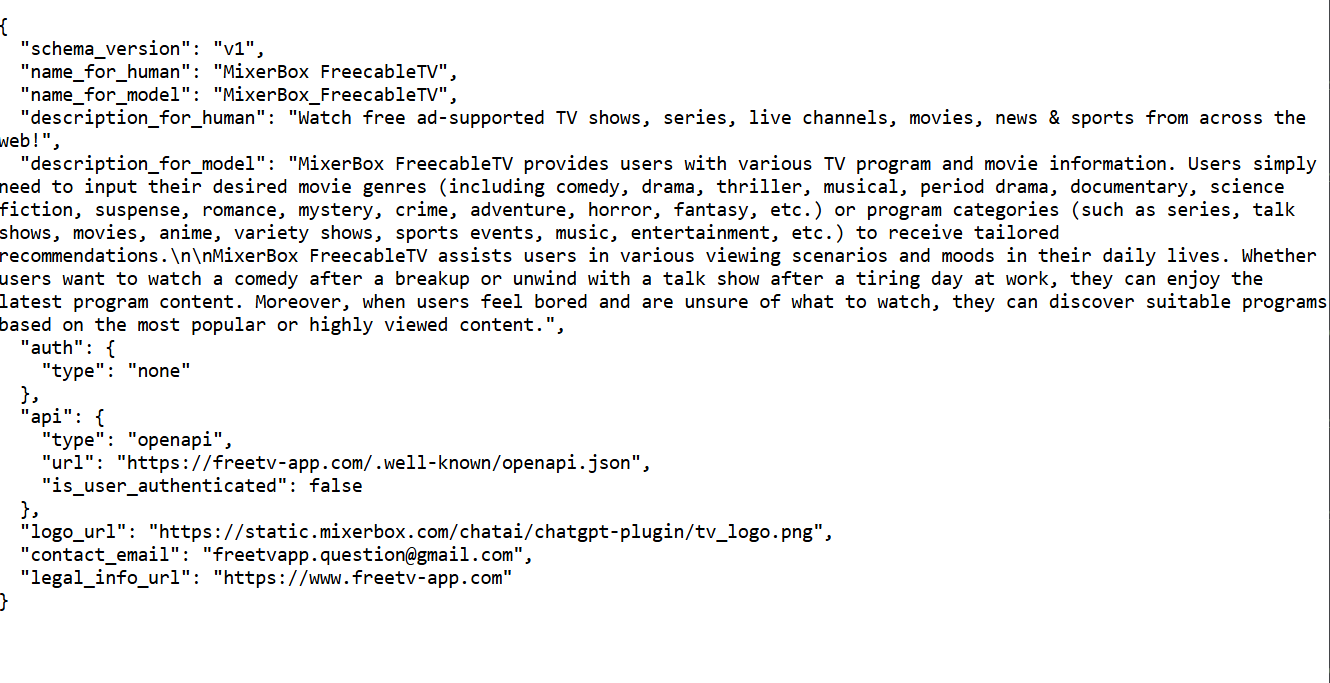}
    \caption{MixerBox shared manifest file}
    \label{fig:enter-label}
\end{figure}

% 在本案例研究中，我们发现 “MixerBox ”品牌下的 18 个插件共享相同的描述和清单文件，这种共享策略不仅带来了用户体验和权限管理方面的问题，还对模型的调用准确性产生了负面影响。

% 首先，由于所有 18 个插件都有相同的 name_for_model 字段，当模型试图调用一个插件的特定功能时，可能会触发多个插件的响应。这可能会导致调用不相关的插件，从而增加模型识别和处理请求的复杂性。例如，当用户希望调用 “MixerBox 计算器 ”插件的功能时，由于 name_for_model 字段（如 “MixerBox ChatPDF ”或 “MixerBox Calendar”）的重复，模型可能会意外触发其他不相关插件的响应。这种混乱的调用机制会导致不准确的响应，并降低用户体验。

% 其次，共享相同的清单文件意味着所有 18 个插件在清单信息、权限设置和法律信息方面也是相同的。这使得平台和用户很难辨别这些插件的实际权限需求。例如，如果其中一个插件需要访问敏感数据，但所有插件都使用相同的权限配置，用户在为其他功能安装插件时可能会无意中授予不必要的权限，从而导致隐私风险。

%这种一致性问题不仅破坏了用户体验，还可能被滥用来提高插件的曝光度。例如，通过在单一manifest 文件中包含所有插件的描述和权限配置，开发者可以在不同的插件中共享流量和排名，使得每个插件在平台中的排序和可见性都得到提升。

In this case study, we found that 17 plugins under the ‘MixerBox’ brand share the same description and manifest files, and this sharing strategy not only brings problems in terms of user experience and privilege management, but also negatively affects the model's invocation accuracy.

Firstly, since all 17 plugins have the same name\_for\_model field, when a model tries to invoke a specific function of a plugin, it may trigger responses from multiple plugins. This can lead to calls to unrelated plugins, which increases the complexity of the model in recognising and processing the request. For example, when a user wishes to invoke the functionality of the ‘MixerBox Calculator’ plugin, the response may be triggered due to the name\_for\_model field (e.g., ‘MixerBox ChatPDF’ or ’ MixerBox Calendar") are duplicated, the model may accidentally trigger responses from other unrelated plugins. This confusing invocation mechanism can lead to inaccurate responses and degrade the user experience.

Second, sharing the same manifest file means that all 18 plugins are also identical in terms of manifest information, permission settings, and legal information. This makes it difficult for platforms and users to discern the actual permission needs of these plugins. 

For example, if one of the plugins requires access to sensitive data, but all of them use the same permission configuration, users may inadvertently grant unnecessary permissions when installing plugins for other functions, leading to privacy risks.

This consistency issue not only disrupts the user experience, but can also be abused to increase the plugin's exposure. For example, by including all plugin descriptions and permission configurations in a single manifest file, developers can share traffic and rankings across plugins, allowing each plugin to be ranked and visible in the platform

% 方法的不足和局限性

% 缺乏上下文理解：虽然自动化脚本可以有效地识别结构上的不一致性，但它缺乏对实际问题的上下文理解。例如，不同插件共享字段未必总是表示问题，但缺少上下文分析可能导致误报。

% 依赖字段的准确性：分析依赖于 name_for_human 和 name_for_model 等字段的准确标记。如果这些字段没有由开发者清晰地标记出实际意图，方法可能会误解数据，从而影响发现结果的可靠性。

%也没写方法评估
%3页
%gpts长什么样
%gpts怎么做的
%gpts的数据变化
%gpts的新问题
%maybe case study 3
\chapter{New GPTs Explorer Periods}

\begin{figure}[!h]
    \centering
    \includegraphics[width=1\linewidth]{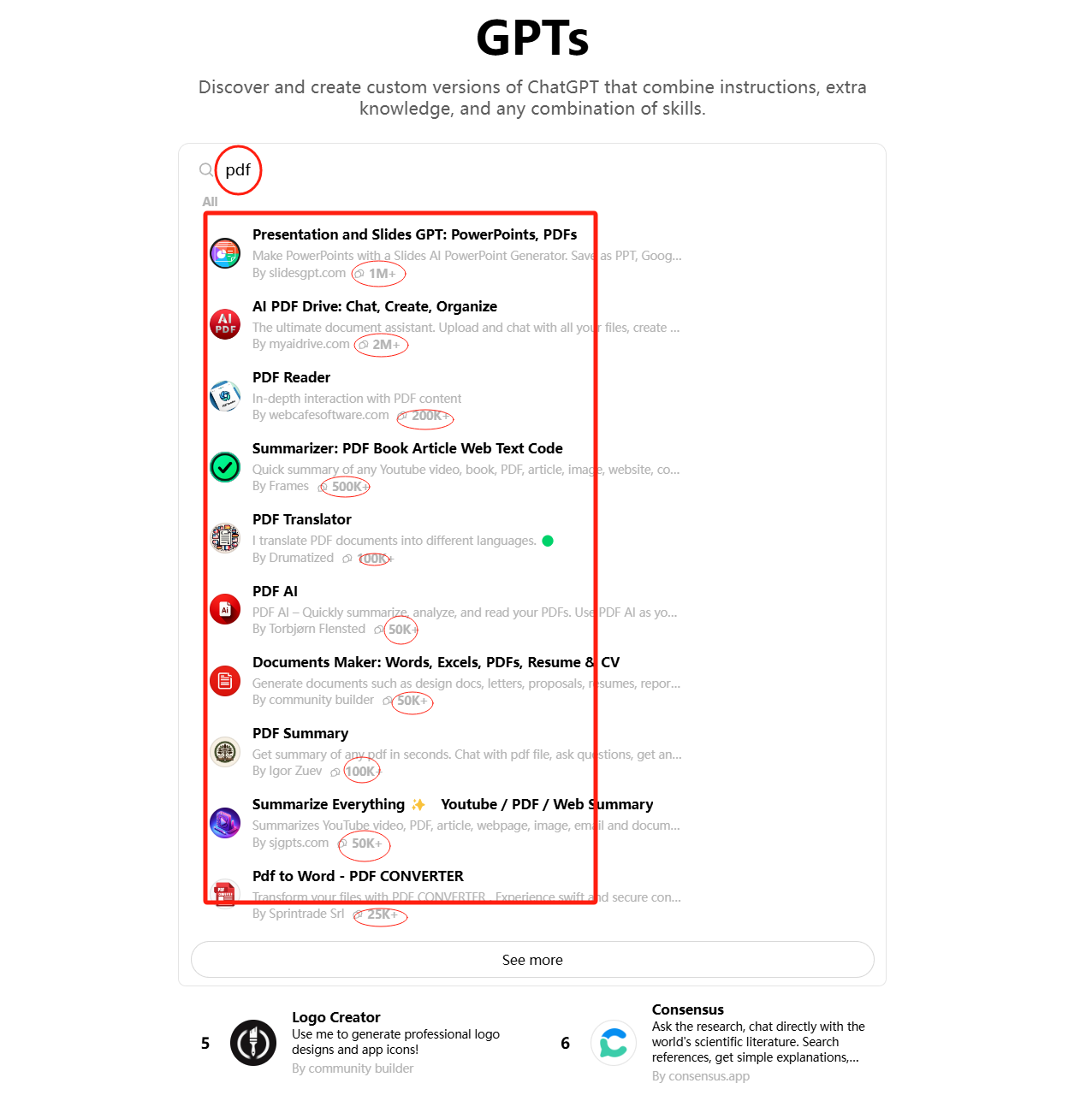}
    \caption{New GPTs Explorer UI}
    \label{fig:enter-label}
\end{figure}
%在2024年4月底，ChatGPT 对插件商店的展示方式进行了更新，推出了新的 GPTs 插件商店 "Explorer GPTs",以更智能的方式呈现。

% 传统商店通过明确的分类和排列展示了所有插件的信息，用户可以浏览商店获得对工具的全貌了解。然而，GPTs 商店通过搜索关键词的方式展示，使得用户难以对所有可用工具有一个完整的认识。这种“隐形内容”的存在，虽有效减少了爬虫，但也可能造成新的如名称盗用等问题。

At the end of April 2024, ChatGPT updated the presentation of the plugin storewith a new GPTs plugin store‘Explorer GPTs’, presented in a smarter way.

The traditional store presents all plugins in a clear categorisation and arrangement, allowing users to browse the store to get a full picture of the tools. However, the GPTs store is presented by search terms, making it difficult for users to get a complete picture of all available tools. The existence of this ‘invisible content’ reduces the number of crawlers, but may also cause new problems such as name theft.

\section{The benefits of new plugin store}
% 首先，新的插件商店更新了UI 展示界面，从列表式展示，更新为搜索引擎式展示，用户必须通过搜索关键词来查找和访问插件，而商店不再提供完整的插件列表。这意味着用户需要准确的关键词来找到最相关的插件，提升了个性化和定向性，帮助用户更高效地发现符合需求的工具。

% 其次，由于新的展示方式，不再采用一览式的排列展示，用户无法浏览商店中的所有插件，而只能依靠关键词获取有限的展示内容。这一机制使得整个商店的插件内容相对“隐形”，用户很难全面了解所有插件的类别和数量。这种“隐藏”设计有效地减少了爬虫的抓取难度，保护了插件信息，同时也在一定程度上限制了用户对插件的认知。

%并且，基于点击量而不是字母顺序排列， 当插件不再按位置排列，插件间的竞争不再集中于列表排名，而是基于关键词的相关性和被引用量，一定程度上解决了恶意竞争和推荐等行为。同时，通过提升相关性，商店能更加精准地将工具推送给真正有需求的用户。

Firstly, the new plugin store updates the UI display interface from a list-based display, to a search engine-based display, where users must search for keywords to find and access plugins, and the store no longer provides a complete list of plugins. This means that users need exact keywords to find the most relevant plugins, improving personalisation and targeting, and helping users discover tools that meet their needs more efficiently.

Secondly, due to the new presentation method, which no longer uses an at-a-glance arrangement of displays, users are unable to browse all plugins in the shop, but can only rely on keywords to access a limited number of displays. This mechanism makes the entire shop's plugin content relatively ‘invisible’, making it difficult for users to get a full picture of the categories and number of plugins. 

This ‘hidden’ design effectively reduces the crawler's difficulty in crawling and protects the plugin information, but also limits the user's knowledge of the plugin to a certain extent.

Moreover, based on the number of clicks rather than alphabetical order, when the plugins are no longer arranged by position, the competition between plugins is no longer focused on the list ranking, but based on the relevance of the keywords and the number of references, which solves the malicious competition and recommendation to a certain extent. At the same time, by improving the relevance, the store can more accurately push the tools to the users who really need them.

However, the recommendation mechanism based on the number of clicks will still cause problems. We should realize that as long as there is an evaluation standard system, the plugin store will inevitably have problems with false reviews and water army manipulation. Such manipulation not only affects user decisions, but also may cover up real security risks\cite{olsson2024fakex}.

\section{New risks of new plugin store}
%尽管新的 GPTs 插件商店带来了显著的优势，但其搜索驱动机制也引发了一些新的风险：
% 名称盗用和品牌模仿的风险：
% 由于搜索结果依赖关键词，这也给了一些开发者利用名称相似性或关键词堆砌来提升自己工具曝光率的机会。例如，某些插件可能会在名称中模仿知名工具，以达到较高的搜索排名。这种名称盗用可能导致用户误选工具，造成不良的用户体验。
% 此外，随着商店中工具数量的增加，这种通过名称或描述模仿的情况会愈发普遍，进而导致用户在选择工具时可能会被误导，难以辨别工具的真实开发者和功能定位。
% 隐形内容与展示透明度的降低：新商店采用搜索驱动的展示方式，导致许多插件内容相对“隐形”，用户难以全面了解商店中的插件种类和数量。限制了用户对插件的认知。
While the new GPTs plugin store brings significant advantages, its search-driven mechanism also raises some new risks:\\

\subsection{Risk of name theft and brand imitation}

Since search results rely on keywords, this also gives some developers the opportunity to exploit name similarity or keyword stacking to boost their tools' exposure. For example, some plugins may mimic well-known tools in their names to achieve higher search rankings. This kind of name theft may cause users to select the tool by mistake, resulting in a poor user experience.

In addition, as the number of tools in the store increases, such imitation by name or description will become more common, which in turn may lead to users being misled in selecting a tool, making it difficult to identify the tool's true developer and functionality positioning.

\subsection{Invisible content and reduced display transparency}
the new store adopts a search-driven display, resulting in many plugins being relatively ‘invisible’, making it difficult for users to fully understand the types and number of plugins in the shop. This limits the user's knowledge of plugins.

\section{Security improvements after reporting}

\begin{table}[!h]
\caption{Comparison of exposures data before and after reporting\label{tab:before_after}}
\small
\resizebox{1\linewidth}{!}{%
\begin{tabular}{c|c c c c c}
\hline

\hline   \multicolumn{1}{c|}{} & \multicolumn{1}{c|}{ \textbf{\makecell[ct]{File\\leakage}} }& \multicolumn{1}{c|}{ \textbf{\makecell[ct]{Inconsistent\\data}} }& \multicolumn{1}{c|}{ \textbf{\makecell[ct]{No\\token}}}& \multicolumn{1}{c|}{ \textbf{\makecell[ct]{Oauth\\auth}}}& \multicolumn{1}{c}{ \textbf{\makecell[ct]{Bearer\\Token}}}\\ \cline{1-6}

  \multicolumn{1}{c|}{\textbf{\makecell[ct]{First assessment}}} & \multicolumn{1}{c|}{368} & \multicolumn{1}{c|}{69} & \multicolumn{1}{c|}{141} & \multicolumn{1}{c|}{27} & \multicolumn{1}{c}{5}\\ \hline
 
 \textbf{Revisit} &\multicolumn{1}{c|}{282} &\multicolumn{1}{c|}{61} &\multicolumn{1}{c|}{89} &\multicolumn{1}{c|}{\textbf{17}}&\multicolumn{1}{c}{3} \\ \hline
 
 \textbf{ Change }&\multicolumn{1}{c|}{\textbf{-23.4\%}} &\multicolumn{1}{c|}{\textbf{-11.6\%}} &\multicolumn{1}{c|}{\textbf{-36.9\%}} &\multicolumn{1}{c|}{\textbf{-37.03\%}}&\multicolumn{1}{c}{\textbf{-40.00\%}} \\ \hline 
 \hline
\end{tabular}}
\vspace{-0.4cm}
\end{table}

% 几乎在同一时间，我们向 OpenAI 报告了上述泄漏问题，并得到了反馈。因此，在 GPTs 系统上线后，我们根据数据集中的插件列表再次进行了测试，并使用相同的脚本验证了之前的反馈。

% 测试结果表明，各个层面的安全漏洞问题都得到了一定程度的缓解。首先，不需要身份验证的插件数量大幅减少，插件的 name\_for\_human 和 name\_for\_model 之间的不一致性也得到了改善，降低了用户误操作和权限滥用的风险。其次，增强了 OAuth 和 Bearer Token 的认证机制，大大减少了 OAuth token 失效的问题。其中，
% 数据不一致性问题得到改善：

% 数据不一致（如插件名称 name_for_human 与 name_for_model 不匹配）从 69 减少到 61，降低了 11.6%。一致性的提升减少了用户对插件功能的误解，并降低了错误选择和权限滥用的风险。

% 文件泄露情况的减少：文件泄露数量从 368 降至 282，减少了 23.4%。文件和资源访问权限控制更严格，降低了因路径或敏感信息暴露导致的安全风险。

% 无认证插件数量减少：通过构造外部 API 请求获取有效结果的无认证插件数量从 141 降至 89，减少了 36.9%。

% OAuth 认证增强：OAuth 认证下的 API 层泄露数量从 27 降至 17，减少了 37%。

% Bearer Token 泄露减少：Bearer Token 下的有效 API 请求数从 5 降至 3，减少了 40%。。

Almost at the same time, we reported the above leak to OpenAI and received feedback. Therefore, after the GPTs system went live, we tested it again against the list of plugins in the dataset and verified the previous feedback using the same script.

The results of the tests showed that the security vulnerability issue was mitigated to some extent at all levels. Firstly, the number of plugins that do not require authentication has been significantly reduced, and the inconsistency between the name\_for\_human and name\_for\_model of the plugins has been improved, reducing the risk of user misuse and privilege abuse.Among them:

\paragraph{Data inconsistencies:}The situation were reduced from 69 to 61(e.g, plugin name\_for\_human does not match name\_for\_model) , a reduction of 11.6\%. The improved consistency reduces user misunderstandings about plugin functionality and reduces the risk of incorrect selection and permission abuse.

\paragraph{Reduction in file leaks:} The number of file leaks decreased from 368 to 282, a 23.4\% reduction. Tighter control of file and resource access permissions reduces security risks due to exposure of paths or sensitive information.

Secondly, the authentication mechanism of OAuth and Bearer Token has been enhanced, greatly reducing the problem of OAuth token failure. Among them:

\paragraph{Reduction in the number of unauthenticated plugins:} The number of unauthenticated plugins that obtain valid results by constructing external API requests has been reduced from 141 to 89, a reduction of 36.9\%.

\paragraph{OAuth Authentication Enhancement:} The number of API layer leaks under OAuth authentication decreased from 27 to 17, a 37\% reduction.

\paragraph{Bearer Token Leakage Reduction:} The number of valid API requests under Bearer Token decreased from 5 to 3, a 40\% reduction.

%仔细写清楚吧就
\chapter{ Potential Risks, Recommendations for Improvement, Limitations, and Future Work}

\section{Potential Risks of Chatgpt plugin store}
% 分析使用 GPTs 商店和插件可能带来的风险，如信息安全、用户隐私、恶意插件的风险等。
%基于以上三层泄露分析的现状，我们分析了chatgpt插件可能造成的风险和危害，以及可能对用户和开发者造成的影响：
% 外部高频攻击：由于manifest文件和API构造方式的泄露，某些插件可能成为高频攻击的目标。攻击者通过自动化工具反复发起请求，试图获取敏感信息或利用系统漏洞进行攻击。这种高频攻击不仅增加了服务器负担，还可能导致服务中断和用户体验下降。

% 未经授权的数据访问：在清单层和API层中，未经过严格身份验证的请求可能导致敏感数据的泄露。攻击者利用系统漏洞，绕过身份验证并获取未经授权的数据访问权限，从而增加数据泄露的风险。这种情况进一步可能导致身份盗窃、滥用权限及其他严重后果，损害用户隐私和安全。

% 恶意插件伪装：某些恶意开发者可能利用插件名称或描述的相似性，误导用户下载和使用恶意插件。鉴于目前交互的不设防，他也可以通过中间层的伪装或者劫持流量，来获取用户和开发者交互信息。这种伪装可能使用户在不知情的情况下授权敏感数据访问，导致数据泄露或滥用。这不仅危害了用户的个人信息安全，也可能对整个插件生态系统的信誉造成损害。

Based on the current state of the three tiers of leakage analysis above, we analyse the risks and hazards that chatgpt plugins may pose, and the impact they may have on users and developers:

\paragraph{External high-frequency attacks:} due to the leakage of manifest files and API construction methods, certain plugins may become the target of high-frequency attacks. Attackers repeatedly launch requests through automated tools in an attempt to obtain sensitive information or exploit system vulnerabilities. Such high-frequency attacks not only increase the burden on the server, but may also lead to service disruption and degradation of user experience.

\paragraph{Unauthorised data access:} In the inventory and API layers, requests without strict authentication can lead to leakage of sensitive data. Attackers exploit system vulnerabilities to bypass authentication and gain unauthorised access to data, thereby increasing the risk of data leakage. This situation may further lead to identity theft, misuse of privileges and other serious consequences that compromise user privacy and security.

\paragraph{Malicious Plugin Disguise:} Certain malicious developers may take advantage of similarities in plugin names or descriptions to mislead users into downloading and using malicious plugins. Given the current undefended nature of the interaction, he may also obtain information about the user-developer interaction by disguising or hijacking traffic in the middle layer. This masquerading may allow users to unknowingly authorise sensitive data access, leading to data leakage or misuse. This not only jeopardises the security of the user's personal information, but may also cause damage to the credibility of the entire plugin ecosystem.

\section{Recommendations for Improvement}
% 提出针对 GPTs 商店的优化建议，例如改进搜索算法、增加插件的透明度等。
%基于泄露问题的现状和目前可能的危害，我们提出如下针对 GPTs 商店的优化建议：
% 加强manifest文件的保护：我们希望平台方对manifest文件的访问权限进行严格控制，确保敏感信息不会因配置错误而暴露，降低潜在攻击面。插件方应该同时控制其可见性和可到达性，并且由于清单托管在第三方网址，应该要求开发者有一定的加密手段。

% 增强身份验证机制：对于插件开发者来说，需要在API层强化身份验证措施，确保所有外部请求都经过适当验证，从而防止未授权访问和数据泄露。

% 提升插件名称和描述的唯一性：在排名和搜索规则上，平台方应该有更加严格的审核，来防止插件的滥用和恶意插件的冒充。

% 实施高频请求监测：建立相应的监测机制，及时识别和响应异常请求模式，防止外部高频攻击。

Based on the current state of the leakage problem and the current possible harms, we propose the following optimisation suggestions for GPTs store:

\paragraph{Strengthen the protection of manifest files:} We hope that platform parties should strictly control the access rights of manifest files to ensure that sensitive information will not be exposed due to misconfiguration and reduce the potential attack surface. The plugin side should control both its visibility and reachability, and since the manifest is hosted on a third-party URL, the developer should be required to have certain encryption means.

\paragraph{Enhance authentication mechanisms:} For plugin developers, authentication measures need to be strengthened at the API layer to ensure that all external requests are properly authenticated, thus preventing unauthorised access and data leakage.

\paragraph{Enhance the uniqueness of plugin names and descriptions:} Platforms should have stricter audits on ranking and search rules to prevent plugin abuse and impersonation of malicious plugins.

\paragraph{Implement high-frequency request monitoring:} Establish appropriate monitoring mechanisms to identify and respond to abnormal request patterns in a timely manner to prevent external high-frequency attacks.

\paragraph{Add corresponding description fields to the plugin manifest file:} Based on other research on plugin stores\cite{ernst2014collaborative}, the manifest file can require mark the data flow and sensitive data processing methods. In this way, the audit system can more effectively evaluate the security of plugins through these fields and improve user security.

\section{Limitations}
% 事实上，本研究方法依旧有相当的局限性，这些局限性影响了API的探查结果及其安全性评估，有些在实际操作中被规避，有点依然存在，包括：

%。首先，在结果的鲁棒性上，常遇到403（禁止访问）和404（找不到资源）等HTTP错误，虽然频繁切换IP和服务器可缓解部分403错误，但对于地域限制的API，这种方法无法完全消除问题。此外，服务器在高并发时也可能出现超时和500错误。这些错误在多次测试中会互相变化，我们无法完全有效分辨，哪些出自于请求发起方的问题，哪些出自于插件方本身限制。

% 其次，在api层，请求体构建的准确性不足也是一大挑战，部分API对请求格式和内容有严格要求，若未完全匹配，可能导致请求失败。我们构建api的手段不够多，这也是导致请求失败的一部分隐藏原因。

% 最后，复杂场景的忽略也是一个问题。特别是api层，多步交互场景可能被忽视，而这些场景需要多个步骤才能完成，单次请求无法捕捉这些交互中的潜在安全隐患。这可能导致对API安全风险的低估。
 In fact, the methodology of this study still has considerable limitations that affect the results of API probing and its security assessment, some of which are circumvented in practice and some of which remain, including:

Firstly, in terms of the robustness of the results, HTTP errors such as 403 (access forbidden) and 404 (resource not found) are often encountered, and although frequent switching of IPs and servers can alleviate some of the 403 errors, this method cannot completely eliminate the problem for geographically restricted APIs. Additionally, timeouts and 500 errors can occur when the server is highly concurrent. These errors vary from one to another over multiple tests, and we can't quite tell which ones are from the request initiator and which ones are from the plugin's own limitations.

Secondly, at the api level, the lack of accuracy in request body construction is also a major challenge. Some APIs have strict requirements on request format and content, which may lead to request failure if they do not match exactly. We don't have enough means to build the api, which is part of the hidden reasons for request failure.

Finally, the ignoring of complex scenarios is also a problem. Especially in the api layer, multi-step interaction scenarios may be ignored which require multiple steps to complete, and a single request cannot capture the potential security risks in these interactions. This may lead to underestimation of API security risks.

\section{Future Work}

%在未来，我们的工作主要集中在两个方面:
%首先，我们将继续追踪和探测新GPT商店中的插件情况，分析是否出现新的问题，以评估其安全性和用户体验。我们需要考虑突破全局爬虫问题，然后是新出现的比如名称冒用等问题。我们将持续跟踪插件的性能，并识别潜在的安全隐患。

% 其次，我们将致力于解决测试框架鲁棒性和智能生成请求体的问题。具体而言，我们计划优化现有的请求构建机制，引入LLM,生成合适的请求主题，以提高请求体的准确性和有效性，我们同时希望深度探究测试框架结果鲁棒性的问题，进一步分析不同返回结果的成因和变化原因，从而使测试框架更合理。

In the future, our work will focus on two aspects:

First, we will continue to track and detect the plugin situation in the new GPTs store, and analyze whether new problems have emerged to evaluate its security and user experience. We need to consider breaking through the global crawler problem, and then new problems such as name fraud. We will continue to track the performance of plugins and identify potential security risks.

Secondly, we will work on solving the problems of test framework robustness and intelligent request body generation. Specifically, we plan to optimize the existing request construction mechanism, introduce LLM, and generate appropriate request subjects to improve the accuracy and effectiveness of the request body. We also hope to deeply explore the problem of robustness of test framework results, and further analyze the causes and changes of different return results, so as to make the test framework more reasonable.

\chapter{Conclusion}

% 在本研究中，我们深入分析了ChatGPT插件在安全性和权限管理方面的潜在风险及其影响。通过对权限泄露、插件一致性等方面的研究，我们识别了多个关键问题，包括未经授权的数据访问、恶意插件伪装以及外部高频攻击等。这些问题不仅可能导致用户敏感信息的泄露，还可能对平台的整体安全性产生负面影响。

% 我们的研究还提出了一系列改进建议，包括加强域名验证、优化API请求构建以及提升插件元数据的一致性等。这些措施旨在减少潜在的安全隐患，提升用户对插件的信任度和使用体验。

% 通过对新GPT商店的追踪和探测，我们将进一步完善安全测试框架，未来的研究将致力于解决鲁棒性和请求体生成的问题，以提高探查的效率和准确性，从而更好地发现和评估问题。

In this study, we deeply analyzed the potential risks and impacts of the ChatGPT plugin in terms of security and permission management. Through research on permission leakage, plugin consistency, etc., we identified several key issues, including unauthorized data access, malicious plugin disguise, and external high-frequency attacks. These issues may not only lead to the leakage of user sensitive information, but also have a negative impact on the overall security of the platform.

Our research also proposed a series of improvement suggestions, including strengthening domain name verification, optimizing API request construction, and improving the consistency of plugin metadata. These measures are aimed at reducing potential security risks and improving users' trust in the plugin and usage experience.

Through tracking and detecting the new GPT store, we will improve the security testing framework. Future research will focus on solving the problems of robustness and request body generation to improve the efficiency and accuracy of exploration, so as to better discover and evaluate problems.

\bibliographystyle{IEEEtran}
\bibliography{references}

\end{document}